# Model-based Pricing for Machine Learning in a Data Marketplace

Lingjiao Chen†, Paraschos Koutris† and Arun Kumar‡

†Department of Computer Science, University of Wisconsin, Madison
‡ Department of Computer Science and Engineering, University of California, San Diego

**Abstract**

Data analytics using machine learning (ML) has become ubiquitous in science, business intelligence, journalism and many other domains. While a lot of work focuses on reducing the training cost, inference runtime and storage cost of ML models, little work studies how to reduce the cost of data acquisition, which potentially leads to a loss of sellers' revenue and buyers' affordability and efficiency. In this paper, we propose a *model-based pricing* (MBP) framework, which instead of pricing the data, directly prices ML model instances. We first formally describe the desired properties of the MBP framework, with a focus on avoiding *arbitrage*. Next, we show a concrete realization of the MBP framework via a noise injection approach, which provably satisfies the desired formal properties. Based on the proposed framework, we then provide algorithmic solutions on how the seller can assign prices to models under different market scenarios (such as to maximize revenue). Finally, we conduct extensive experiments, which validate that the MBP framework can provide high revenue to the seller, high affordability to the buyer, and also operate on low runtime cost.

## 1 Introduction

Data analytics using machine learning (ML) is an integral part of science, business intelligence, journalism, and many other domains. Research and industrial efforts have largely focused on performance, scalability and integration of ML with data management systems [11, 23, 30]. However, limited research has studied the *cost of acquiring data* for ML-based analytics.

Users often buy rich structured (*relational*) data to train their ML models, either directly through companies (e.g., Bloomberg, Twitter), or through *data markets* (e.g., BDEX [1],Qlik [2]). Such datasets are often very expensive due to the immense effort that goes into collecting, integrating, and cleaning them. Existing pricing schemes either force users to buy the whole dataset or support simplistic pricing mechanisms, without any awareness of ML tasks (e.g., the dataset is typically used to train predictive models). This means that valuable datasets may not be affordable to potential buyers with limited budgets, and also that data sellers operate in inefficient markets, where they do not maximize their potential revenue. Simplistic pricing schemes may also create undesirable arbitrage opportunities. Thus, as [12] also points out, *there is a need to transition from markets that sell only data to markets that can also directly sell ML models.*

***Model-based Pricing.*** In this paper, we take a first step towards the long-term vision of creating a marketplace for selling and buying ML models, by presenting *a formal and practical fine-grained pricing framework for machine learning over relational data.* Our key observation is that, instead of selling raw data to the buyers, it makes more sense to directly sell *ML model instances* with different accuracy options. The price then should depend on the accuracy of the model purchased, and not the underlying dataset. Since the price is based on the model instance, we call our framework *model-based pricing* (MBP).

The high level view of MBP is demonstrated in Figure 1. The data market involves three agents, namely, the *seller* who provides the datasets, the *buyer* who is interested in buying ML model instances, and the *broker* (market) who interacts between the seller and the buyer. First, the seller and/or the broker perform market research to ascertain curves representing demand and value for the ML model instances among potential buyers. These curves plot demand and value as a function of the error/accuracy of the ML model trained. The broker uses the market research information to build price-error curves that are presented to the buyers. The buyer specifies a desired price or error budget and pays the broker, who computes an appropriate ML model instance (according to the buyer's specifications), and returns it to the buyer. We should note here that the broker provides different price-error curves, depending on the ML model that the buyer desires.

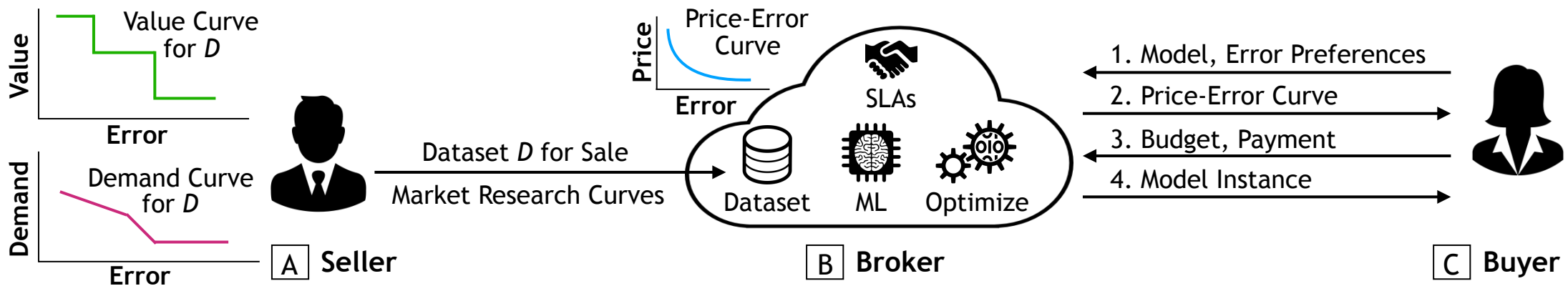

Figure 1: Model-based pricing market setup. (A) The *seller* is the agent who wants to sell ML model instances trained on their commercially valuable dataset $D$. (B) The *broker* is the agent that mediates the sale for a set of supported ML models and gets a cut from the seller for each sale. (C) The *buyer* is the agent interested in buying a ML model instance trained on $D$.

***Desiderata and Challenges.*** Achieving the MBP framework is a technically challenging task, from both theoretical and practical points of view. First, in order to guarantee *affordability*, the MBP framework must allow buyers with different budgets to buy model instances with different accuracy guarantees. However, the model instance generation should be performed *efficiently* with low runtime cost, since model training is typically time-consuming. Second, the MBP framework must prevent *arbitrage* opportunities, where a buyer can combine model instances of low accuracy and low price to obtain a high accuracy instance for a cheaper price than the one provided by the market. For example, an instance with high accuracy should always be at least as expensive as an instance with lower accuracy. Finally, the MBP framework must provide capabilities for the broker/buyer to assign the prices such that the revenue is maximized.

***Our Solution.*** Our key technical contribution is a simple and efficient *noise-injection mechanism* that realizes an MBP framework with formal guarantees. Specifically, the broker first trains the optimal model instance, which is a one-time cost. When a buyer requests a model instance, the broker adds random Gaussian noise to the optimal model and returns it to the buyer. Our proposed mechanism avoids training a model instance from scratch and is able to achieve real time interaction. We show that the error of the ML model instance (when certain properties are satisfied by the error function) is a monotone function of the *variance* of the noise injected in the model. Hence, the variance works as a parameter that controls the magnitude of the error.

The pricing mechanism charges a price according to the variance of the noise injected to the model instance. Adding noise with low variance implies a model instance with expected low error and thus high price, while noise with high variance results in an instance with expected larger error and low price. This enables the buyer to either choose cheaper but less accurate instances or more accurate yet more expensive ones. Essentially, our mechanism provides different *versions* of the desired ML model of varying quality, in analogy to the notion of versioning in information selling [29].

Our proposed MBP mechanism comes with a concise characterization of when a pricing function is provably arbitrage-free. In the main theoretical result of this paper, we show that a pricing function is arbitrage-free if and only if the price of a (randomized) model instance is monotone increasing and subadditive with respect to the inverse of the variance. For example, this means that when we double the variance, we should at most cut the price in half, otherwise we would create an arbitrage opportunity.

For revenue maximization, we establish a formal optimization framework based on the buyer's value and demand curves. We show that revenue maximization is a computationally hard problem even under a simple revenue model. To address this intractability, we present a novel method of relaxing the subadditive constraints, which allows us to obtain polynomial time algorithms with provable approximation guarantees. Central to the revenue maximization problem in our setting is the problem of interpolating a monotone and subadditive function through given points, which could be of independent interest.
Finally, we prototype the MBP framework in standalone MATLAB, which is popular for ML-based analytics (but note that our framework is generic and applicable in other settings as well). We present an extensive empirical evaluation using both real and synthetic datasets. Our experiments validate that MBP always attains the highest revenue and provides the highest buyer affordability compared to the existing naive solutions, while simultaneously guaranteeing protection against arbitrage. We also show that our revenue maximization solution for price setting is orders of magnitude faster than brute-force search, while empirically achieving a revenue with negligible gap to the optimal revenue.

***Summary of Contributions.*** In summary, this paper makes the following contributions.

- To the best of our knowledge, this is the first paper on a formal framework of ML-aware model-based pricing. We identify and formally characterize important properties, such as arbitrage freeness, that such a framework should satisfy.
- We propose a concrete MBP mechanism via an noise injection approach, and establish a concise characterization of the desired properties of a pricing function.
- We develop a revenue optimization framework that finds the optimal prices with the desired properties as constraints. Although it is a provably computationally hard problem, we provide an approximate solution which gives a pricing function with provably high revenue.
- Finally, extensive experiments validate that our proposed solution provides high revenue for the seller, large affordability ratio for the buyer, and fast runtime for the broker.

***Outline.*** Section 2 presents the problem setup and background. Section 3 introduces the MBP framework and relevant desiderata. Section 4 provides a concrete realization of the MBP framework via the noise injection approach. Section 5 studies the revenue optimization problem and Section 6 presents the implementation and experimental evaluation. We conclude in Section 7. All missing proof is left to the appendix .

## 2 Related Work

In this section, we discuss related work on data pricing and machine learning.

***Pricing Relational Queries.*** The problem of pricing relational data has received a lot of attention recently. The pricing schemes currently supported in data markets are typically simplistic: a prospective buyer can either buy the whole dataset for some fixed price, or ask simple queries and be priced according to the number of output tuples. A recent line of work [4,13–15,20,22,28] has formally studied pricing schemes for assigning prices to relational queries issued over such data, a setting called *query-based pricing*. In query-based pricing, we are given a dataset $D$ and a query $Q$, and the goal is to assign a price $p(Q, D)$ based on the information disclosed by the query answer. Central to query-based pricing is the notion of *arbitrage*. Intuitively, whenever query $Q_1$ discloses more information than query $Q_2$, we want to ensure that $p(Q_1) \geq p(Q_2)$; otherwise, the data buyer has an arbitrage opportunity to purchase the desired information for a lesser price. To formalize information disclosure, query-based pricing uses the mathematical tool of *query determinacy* [25, 26]. In the proposed framework [13], the seller specifies a set of fixed prices for views over the data (*price points*), based on which an arbitrage-free price is computed for any query. [22] provides several necessary conditions for arbitrage-free pricing functions. [19] takes a first step towards pricing private datasets.
At first glance, MBP seems similar to *query-based pricing*. For relational queries, the price is for the information released by the query output, while for ML analytics, the price is for the information released by the model instance. However, there are fundamental differences: for relational queries, the buyer obtains a *deterministic* answer, while for ML analytics, the model is typically computed in an *non-deterministic* way. Also, in MBP, we enable the buyer to specify an *accuracy* constraint to control the predictive power of the model instance they buy. Our MBP mechanism is closer to that of [19], where Laplacian noise is added to the result of aggregates to protect individuals from privacy loss.

***Markets for ML.*** While quite a few ML systems [3, 5, 6, 8, 21, 27] have been developed to reduce the computational cost of training ML models, there has been little attention to the problem of constructing ML markets until recently [12, 17]. [17] develops a market system via block chain technology to allow exchange and purchasing of ML models. [12] points out the importance of creating markets for ML applications. MBP can be viewed as the first foray into how we should build such markets.

***ML over Relational Data.*** We focus on standard supervised ML for relational/structured data, specifically, classification and regression. We are given a dataset table $D$ with $n$ labeled examples and $d$ features. The target (label) is denoted $Y$, while the feature vector is denoted $X$. We assume that **X** and $Y$ correspond to the *attributes* of a single relation. In this setting, the labeled examples are typically assumed to independently and identically distributed (IID) samples from some underlying (hidden) distribution that produces the data, $P[\mathbf{X}, Y]$. An *ML model* is simply a mathematical model to approximate $P$ in some intuitive manner. For example, the least squares linear regression model assumes the data can be represented using $d$-dimensional hyperplane. An *ML model instance* is a specific instance of that ML model that corresponds to some prediction function $f : \mathcal{D}_{\mathbf{X}} \to \mathcal{D}_Y$. For example, an instance of the least squares linear regression model a given vector $w \in \mathbb{R}^d$. Given $D$, a *learning algorithm* computes such a prediction function. The set of functions learnable (representable) by an ML model is called its *hypothesis space*. The predictive power of a model instance is often *evaluated* using standard scores such as *holdout test error* [9]. There are hundreds of ML models [24]; some of the most popular ones are Naive Bayes, other Bayesian Networks, and Generalized Linear Models (GLMs). These models are popular mainly due to their *interpretability*, simplicity, speed, and extensive systems support. Thus, we primarily focus on such models.

# 3 Model-based Pricing Framework

In this section, we introduce the framework of model-based pricing (MBP), and then outline some basic properties that our framework must satisfy. We summarize the notations used throughout this paper in Table 1.

## 3.1 Market Setup and Agents

Our framework involves three types of agents that interact in the setting of a data marketplace: the *seller*, the *broker* and the *buyer*. We now introduce our market setup involving these agents and their interactions, as well as the notation and assumptions we use. Figure 1 illustrates the market setup.

***Seller.*** The seller provides the dataset $D$ for sale, and it is given as a pair $(D_{\text{train}}, D_{\text{test}})$, wherein $D_{train}$ is called the *train set* (for obtaining model instances) and $D_{test}$ is the *test set*. This train-test split is standard in ML practice [9]. For simplicity of exposition, we express a row in $D$ as a labeled example of the form $z = (\mathbf{x}, y)$, where $\mathbf{x} = z[\mathbf{X}]$ is the feature vector and $y = z[Y]$ is the target.

***Broker.*** The broker specifies a menu of ML models $\mathcal{M}$ she can support (*e.g.*, logistic regression for classification and ordinary least squares for regression), along with the corresponding hypothesis spaces $\mathcal{H}_m$ for each $m \in \mathcal{M}$. For now, fix an ML model, *i.e.*, the hypothesis space $\mathcal{H}$. An *error (loss) function* $\lambda(h, D)$ measures the goodness of a hypothesis $h \in \mathcal{H}$ on $D_{\text{train}}$ and returns a real number in $[0, \infty)$. Given $D$ and the error function $\lambda$, let $h^*_\lambda(D) = \arg\min_{h \in \mathcal{H}} \lambda(h, D)$ denote the optimal model instance, *i.e.*, the model instance that obtains the smallest error on the training dataset w.r.t. $\lambda$. We also define another error function $\epsilon$ that can operate on either $D_{\text{test}}$ or $D_{\text{train}}$, based on the buyer's preference. For simplicity of exposition, we use $D$ with both the error functions, with the implicit convention that $\lambda$ operates on $D_{\text{train}}$ and $\epsilon$ on $D_{\text{test}}$ or $D_{\text{train}}$.

Table 1: Notations.

| Symbol | Meaning |
|---|---|
| $D/D_{\text{train}}/D_{\text{test}}$ | dataset/train set/test set |
| $n_0/n_1/n_2$ | number of samples in $D/D_{\text{train}}/D_{\text{test}}$ |
| $d$ | number of features |
| $\mathbf{x}/y$ | feature vector/target value (label) |
| $\mathcal{M}/m$ | a set of ML models/a specific model |
| $\mathcal{H}/h$ | hypothesis space/a specific hypothesis |
| $\lambda(\cdot,\cdot)/\epsilon(\cdot,\cdot)$ | error function for training/accuracy report |
| $h^*_\lambda(D)$ | optimal model instance w.r.t. $\lambda$ on $D$ |
| $\delta$ | noise control parameter (NCP) |
| $W_\delta$ | distribution generated by $\delta$ |
| $w \sim W_\delta$ | random variable generated by $W_\delta$ |
| $\mathcal{K}(\cdot,\cdot)$ | randomized noise mechanism |
| $\hat{h}^\delta_\lambda(D) = \mathcal{K}(h^*_\lambda(D), w)$ | model generated via $\mathcal{K}$ |
| $p_{\epsilon,\lambda}(\delta, D)$ | price of the model instance $\hat{h}^\delta_\lambda(D)$ |
| $\bar{p}(x) = p_{\epsilon,\lambda}(1/x, D)$ | price of the model instance $\hat{h}^{1/x}_\lambda(D)$ |

In general, $\epsilon$ can be different from $\lambda$ because that may be more meaningful from an ML accuracy standpoint. In particular, in this paper, we focus on the following types of ML models and their associated error functions. Formally, we focus on $\lambda$ that is *strictly convex*. In particular, for classification, this covers the popular logistic regression and linear SVM model (with L2 regularization). For regression with a real-valued target, this covers the popular least squares linear regression model. We think it is reasonable to focus on these well-understood ML models and leave more complex ML models and error functions to future work, since it enables us to study the issues of model-based pricing in depth in this first paper. However, we emphasize that our market setup, our analyses of the properties of the pricing functions, and the revenue optimization are all generic and applicable to *any* ML model. For $\epsilon$, we use both the same loss function as $\lambda$ and the commonly used *misclassification rate* error function for classification models. We tabulate the ML models and their associated error functions in Table 2.

| **ML model** | **Error Function(s)** |
|---|---|
| For $\lambda$; $(y, \mathbf{x})$ from train set $D_{\text{train}}$ | |
| Lin. reg. | $\sum_{(y,\mathbf{x})}(y - \mathbf{w}^T\mathbf{x})^2$ $[+\mu\|w\|^2]$ |
| Log. reg. | $\sum_{(y,\mathbf{x})} log(1 + e^{-y\mathbf{w}^T\mathbf{x}})$ $[+\mu\|w\|^2]$ |
| L2 Lin. SVM | $\sum_{(y,\mathbf{x})} max(1, -y\mathbf{w}^T\mathbf{x}) + \mu\|w\|^2$ |
| For $\epsilon$; $(y, \mathbf{x})$ from test set $D_{\text{test}}$ or train set $D_{\text{train}}$ | |
| Lin. reg. | Same as $\lambda$ |
| Log. reg.<br>L2 Lin. SVM | Same as $\lambda$; $\sum_{(y,\mathbf{x})} \mathbb{1}_{y=(\mathbf{w}^T\mathbf{x}>0)}$ |

Table 2: ML models in $\mathcal{M}$ and associated error functions. $[\cdot]$ indicates optional regularization. All errors are typically averaged by the number of examples used.

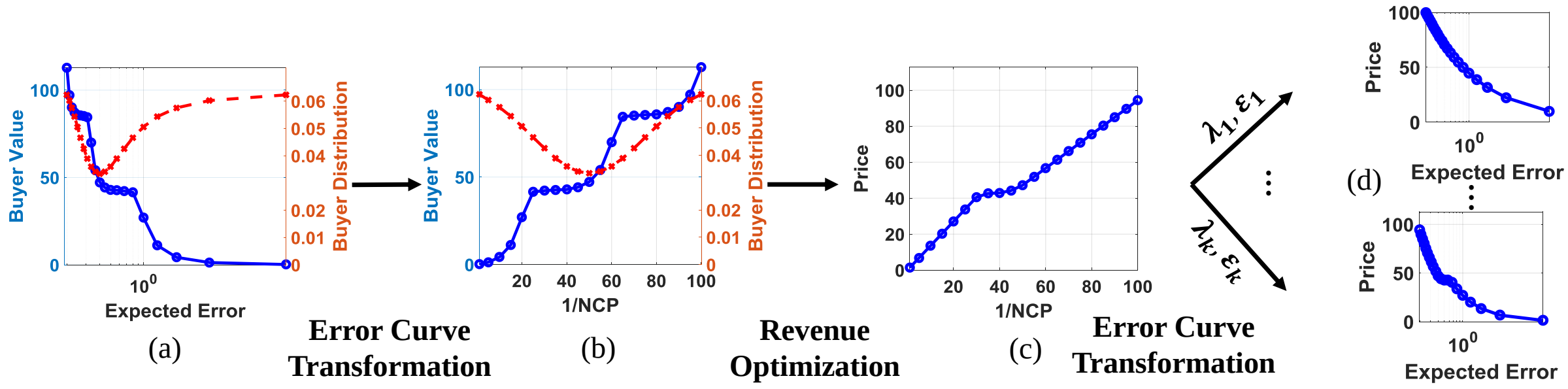

Figure 2: End-to-end model based pricing. The seller first provides the broker with the buyer value and demand curve via market research. The broker then obtains the curve v.r.t the inverse NCP via some error transformation, computes the pricing function via revenue optimization, and then returns a pricing curve to the buyers based on different error functions $\lambda, \epsilon$.

The broker releases a model instance through a *randomized mechanism* $\mathcal{K}$ that enables them to *trade off ML error for the price* the model instance is sold for. This is a key novel technical capability in our market setup that enables model-based pricing in contrast to flat pricing. This mechanism enables us to realize *versioning* in the context of ML, in analogy to the versioning of digital information goods in micro-economics literature [29].

Specifically, $\mathcal{K}$ uses a set of parametrized probability distributions $\{\mathcal{W}_\delta \mid \delta \in \mathbb{R}_+\}$. Given a dataset $D$, an error function $\lambda$ and a noise control parameter (NCP) $\delta$, the broker first computes the optimal model instance $h^*_\lambda(D)$. Then, they sample $w \sim \mathcal{W}_\delta$ and output a *noisy version* of the optimal model, $\hat{h}^\delta_\lambda(D) = \mathcal{K}(h^*_\lambda(D), w)$. The NCP $\delta$ will be used as a knob to control the amount of noise added, and in turn, the price of the model instance sold. We will discuss more about the noise addition mechanism's desiderata shortly, but first, we explain the other agent in our setup.

***Buyer.*** The buyer specifies an ML model $m \in \mathcal{M}$ they are interested in learning over $D$, along with their preferences for the error functions $\lambda$ and $\epsilon$ to use from among the ones the broker supports. After a set of interactions with the broker, which will be explained shortly, the buyer obtains an instance of $m$ that satisfies their price and/or error constraints.

## 3.2 Agent Interaction Models

Having introduced the three agents in our framework, we now explain how the market operates. Figure 1 illustrates our market setup and the interactions between the seller and broker, as well as between the broker and the buyer.

***Broker-Seller Interaction Model.*** Apart from providing $D$, the seller works with the broker to determine the *pricing function* $p$ to use for a given ML model. The pricing function does *not* depend solely on the released model instance $\hat{h}^\delta_\lambda(D)$. Instead, it depends on $D$, the NCP $\delta$, and the two error functions $\lambda, \epsilon$. Hence, we express the pricing function as $p_{\epsilon,\lambda}(\delta, D)$, which returns a non-negative real number in $[0, \infty)$. The desirable properties of a pricing function, how to set them to maximize revenue for the seller but still satisfy potential customers and run a feasible market, and how to compute them efficiently are all core technical challenges that we address later in this paper.

In the context of the interaction model, the broker is able to set the pricing functions based on two curves provided by the seller based on their *market research* about $D$. These curves, illustrated in Figure 2(a), tell the broker how much value potential customers attach to model errors in terms of monetary worth (value curve) and how much demand there is the market for different model errors (demand curve). Defining and using these curves as inputs for optimizing pricing is standard practice in micro-economics for data markets such as the sale of digital videos [10]. Our work adopts this practice for the novel scenario of selling ML model instances trained on $D$.
Given the demand and value curves as a function of some error function, the broker first transforms them to demand and value curves as a function of the inverse NCP, as shown in Figure 2(b). Then, the broker computes the revenue maximizing pricing function as a function of the inverse NCP (Figure 2(c)) – we defer discussion of the revenue optimization problem till Section 5.

***Broker-Buyer Interaction Model.*** The buyer-broker interaction has 4 four steps, as illustrated by Figure 1(C). First, the buyer specifies the ML model they are interested in ($\mathcal{H}$) and the two error functions $\lambda, \epsilon$ corresponding to that model that the broker supports. For instance, these could be the log loss for logistic regression training but the zero-one classification error for testing. Second, given these inputs, the broker computes a curve that plots the price together with the *expected error* for every NCP $\delta$, given by $\mathbb{E}_{\sim \mathcal{W}_\delta}\left[\epsilon\left(\hat{h}^{\delta}_{\lambda}(D), D\right)\right]$. This curve shows to the buyer the possible price points of the different versions of this model in $D$. As shown in Figure 2 (d), different $\lambda, \epsilon$ corresponds to different pricing curves.
For the third step, the buyer has three options. First, she can specify a particular point on the curve (*i.e.* a price-error combination); since we know that $\delta$ behaves monotonically w.r.t. the expected error, the broker can find the unique $\delta^*$ that corresponds to that point, and obtains $\hat{h}^{\delta^*}_{\lambda}(D)$. The second option is that the buyer specifies an *error budget* $\hat{\epsilon}$. The broker then solves the following optimization problem:

$$\delta^* = \arg\min_{\delta} p_{\epsilon,\lambda}(\delta, D)$$
$$\text{s.t.} \quad \mathbb{E}_{\sim \mathcal{W}_\delta}\left[\epsilon\left(\hat{h}^{\delta}_{\lambda}(D), D\right)\right] \leq \hat{\epsilon}$$

The third and final option for the buyer is to specify a *price budget* $\hat{p}$ to the broker. The broker then solves the following optimization problem:

$$\delta^* = \arg\min_{\delta} \mathbb{E}_{\sim \mathcal{W}_\delta}\left[\epsilon\left(\hat{h}^{\delta}_{\lambda}(D), D\right)\right]$$
$$\text{s.t.} \quad p_{\epsilon}(\delta, D) \leq \hat{p}$$

The third step is for the buyer to pay the price of $p_{\epsilon,\lambda}(\delta^*, D)$ to the broker. In the final step of this interaction, the broker gives the obtained model instance $\hat{h}^{\delta^*}_{\lambda}(D)$ to the buyer.

***Restrictions on the Randomized Mechanism.*** The mechanism $\mathcal{K}$ used by the broker needs to satisfy certain properties to enable us to reason about the market's behavior and ensure it is "well-behaved" (a property we will define shortly). In particular, in this paper, we only consider randomized mechanisms that satisfy the following two conditions.

- $\mathcal{K}$ is *unbiased*, which means that:

$$\mathbb{E}_{w \sim \mathcal{W}_\delta}\left[\mathcal{K}(h^*_{\lambda}(D), w)\right] = h^*_{\lambda}(D)$$

  In simple terms, the model instance sold after adding noise is the same as the optimal model instance *in expectation*. Only the NCP $\delta$ controls how much noise is added, and thus, how much degradation there is in the model instance's parameters.
- The parameter $\delta$ behaves monotonically w.r.t. the expected error,

$$\delta_1 \leq \delta_2 \Leftrightarrow \mathbb{E}\left[\epsilon\left(\hat{h}^{\delta_1}_{\lambda}(D), D\right)\right] \leq \mathbb{E}\left[\epsilon\left(\hat{h}^{\delta_2}_{\lambda}(D), D\right)\right]$$

  That is, by increasing the NCP $\delta$ we strictly increase the expected error as well, and vice versa.
  The feasibility of this assumption depends on the exact $\epsilon$ used. As we show later, this assumption is reasonable for many common scenarios and lets us provide formal guarantees on the market's behavior .

We now present two concrete examples of how our model-based pricing framework operates. The first example is for computing a simple SQL-style aggregate, *average*, to make it easier to understand the concepts. The second example is for a common statistical ML model, *linear regression*. We use these two as running examples in the rest of the paper.

**Example 1.** *Suppose the buyer is interested in "learning" the average value of a particular feature (column) of $D$. The hypothesis space $\mathcal{H}$ is just $\mathbb{R}$. The error functions can be simply defined as $\lambda(h, D) = (h - \bar{x})^2$, where $\bar{x}$ is the true column average from $D_{train}$, and similarly for $\epsilon$ on $D_{train}$. One possible randomized mechanism for adding noise is $\mathcal{K}_1(h^*_\lambda(D), w) = h^*_\lambda(D) + w$, where $w \sim U[-\delta, \delta]$, i.e., a uniform random distribution. Yet another possible mechanism is $\mathcal{K}_2(h^*_\lambda(D), w) = h^*_\lambda(D) \cdot w$, where $w \sim U[1 - \delta, 1 + \delta]$. Both of these randomized mechanisms satisfy the two restrictions listed earlier.*

**Example 2.** *Suppose the buyer is interested in learning a least squares linear regression model on $D$. The hypothesis space $\mathcal{H}$ is then the set of all hyperplanes $h \in \mathbb{R}^d$. The error function $\lambda$ is the least squares loss defined on the training subset, i.e.,*

$$\lambda(h, D) = \frac{1}{2|D_{train}|} \sum_{z_i \in D_{train}} \left(h^T \mathbf{x}_i - y_i\right)^2 .$$

*The error function $\epsilon$ is the same as above, except on the test subset $D_{test}$ of $D$. Given the optimal model instance $h^*_\lambda(D)$, one randomized mechanism for adding noise is as follows. Let $\mathcal{W}_\delta = \mathcal{N}(0, \delta^2)$ be the standard d-dimensional normal (Gaussian) distribution with mean 0 and variance $\delta^2$. The noise addition mechanism is as follows:*

$$\mathcal{K}(h^*_\epsilon(D), w) = h^*_\epsilon(D) + w$$

*Thus, we simply add Gaussian noise (of different magnitudes) independently to each co-efficient of the optimal model instance and return it to the buyer. Another possible mechanism is to sample noise from a zero-mean Laplace distribution. Once again, of these randomized mechanisms satisfy the two restrictions listed earlier.*

### 3.3 Pricing Function Desiderata

We now return to the concept of the pricing functions mentioned earlier when explaining the broker-seller interaction model. For the market to be able to work, these pricing functions need to satisfy a set of desiderata that provide some guarantees to both the seller and the buyer. In a sense, these guarantees act as the service-level agreement (SLA) for model-based pricing. In particular, we want the pricing functions to satisfy the following requirements.

***Non-negativity.*** Clearly, the pricing function has to be *non-negative*, since the buyer should not be able to make money from the broker by obtaining an ML model instance.

**Definition 1.** *A pricing function $p_{\epsilon,\lambda}$ is* non-negative *in dataset $D$ iff for every parameter $\delta$ (of $\mathcal{K}$),*

$$p_{\epsilon,\lambda}(\delta, D) \geq 0.$$

***Error Monotonicity.*** Next, we want to make sure that if for a parameter $\delta_1$, we obtain a smaller (or equal) expected error than for a parameter $\delta_2$, then the price is larger (or equal) for the former model instance. Otherwise, a buyer that wants to buy a model instance with the smaller error can purchase it for a smaller price. This situation is illustrated in Figure 3. The formal definition is as follows.

**Definition 2.** *A pricing function $p_{\epsilon,\lambda}$ is* error-monotone *in dataset $D$ if for every parameters $\delta_1, \delta_2$,*

$$\mathbb{E}\left[\epsilon(\hat{h}^{\delta_1}_\lambda(D), D)\right] \leq \mathbb{E}\left[\epsilon(\hat{h}^{\delta_2}_\lambda(D), D)\right]$$

*implies that*

$$p_{\epsilon,\lambda}(\delta_1, D) \geq p_{\epsilon,\lambda}(\delta_2, D).$$

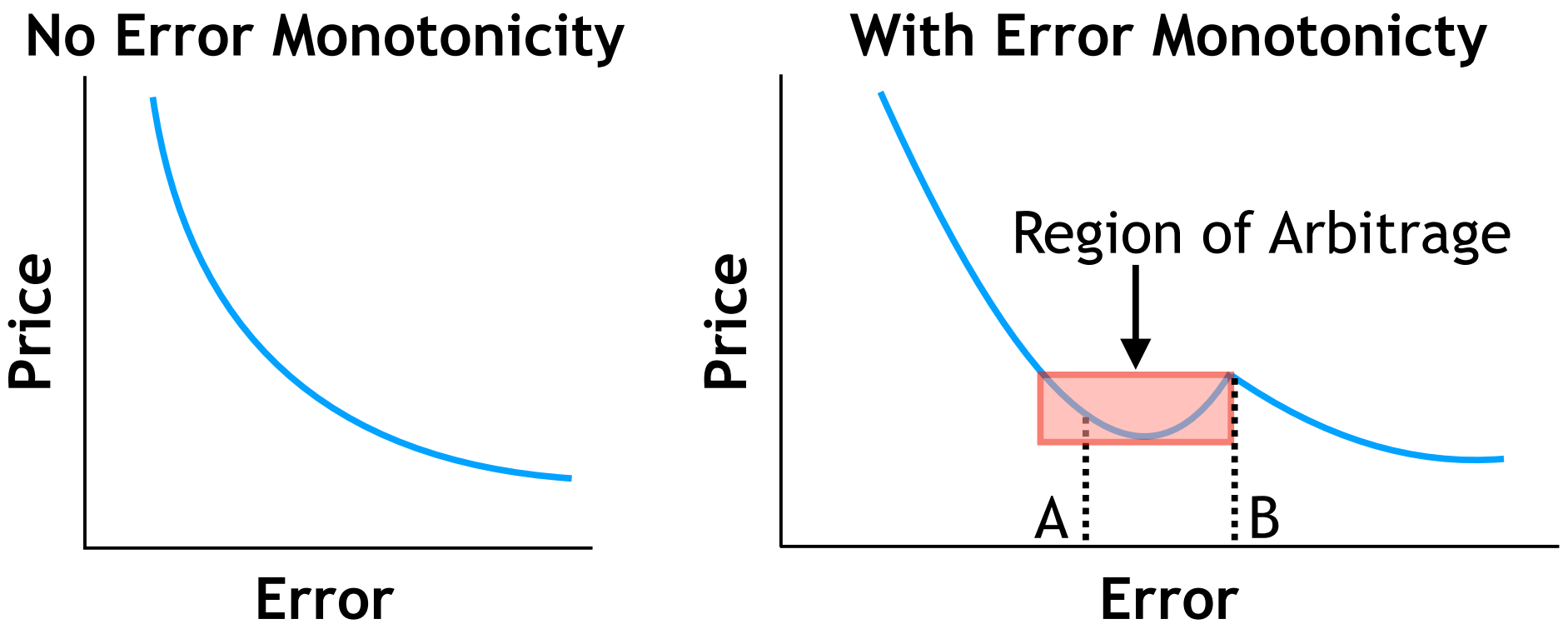

Figure 3: The pricing function on the left has error monotonicity. The one on the right does not, which leads to *arbitrage* situations. Point A has both a lower price and lower error than B. Thus, buyers are unlikely to pick B; indeed, the entire shaded region shown is useless for the seller, since they lose some potential revenue.

Error monotonicity implies that whenever we have two parameters $\delta_1, \delta_2$ such that $\mathbb{E}\left[\epsilon(\hat{h}_\lambda^{\delta_1}(D), D)\right] = \mathbb{E}\left[\epsilon(\hat{h}_\lambda^{\delta_2}(D), D)\right]$, then the prices must be equal as well, *i.e.* $p_{\epsilon,\lambda}(\delta_1, D) = p_{\epsilon,\lambda}(\delta_2, D)$. Hence, the error monotonicity property implies that the price does not depend on the actual parameter $\delta$ of the mechanism, but on the error that this parameter induces.

***Arbitrage-freeness.*** The final property we discuss is *arbitrage-freeness*, which is analogous to a similar notion in query-based pricing [13]. We first explain the importance of this property intuitively. Suppose a buyer wants to buy one model instance with a small error but large price. Suppose further she also buys more of such model instances at different prices, the sum of all of which is lower than that of the desired single model instance. At the same time, suppose she is able to "combine" the latter set of model instances to construct a new model instance with an error smaller than the originally desired single model instance. In this case, she would rather just buy the latter set of model instances instead of the original model instance to get an error lower than what the market is set up for. Such a situation is called *arbitrage*. For the market to work well, we need to ensure that it is *arbitrage-free*, i.e., situations such as these do not happen (or are extremely unlikely). This intuition is captured formally by the following definitions.

**Definition 3** ($k$-Arbitrage). *We say that a pricing function $p_{\epsilon,\lambda}$ exhibits $k$-arbitrage in dataset $D$ if there exist parameters $\delta_0, \delta_1, \delta_2, \cdots, \delta_k$, and a function $g : \mathcal{H}^k \to \mathcal{H}$ such that*

1. $\sum_{i=1}^{k} p_{\epsilon,\lambda}(\delta_i, D) < p_{\epsilon,\lambda}(\delta_0, D)$, *and*
2. $\mathbb{E}\left[\epsilon(\tilde{h}, D)\right] \leq \mathbb{E}\left[\epsilon(\hat{h}_\lambda^{\delta_0}(D), D)\right]$, *where $\tilde{h}$ is the model $\tilde{h} = g(\hat{h}_\lambda^{\delta_1}(D), \hat{h}_\lambda^{\delta_2}(D), \ldots, \hat{h}_\lambda^{\delta_k}(D))$ s.t.* $\mathbb{E}\left[\tilde{h}\right] = h_\lambda^*(D)$.

**Definition 4** (Arbitrage-free). *A pricing function $p_{\epsilon,\lambda}$ is* arbitrage-free *in dataset $D$ iff it does not exhibit $k$-arbitrage for any $k \in \mathbb{N}^+$.*

Not surprisingly, arbitrage-freeness implies that the pricing function is also error monotone, as the following simple lemma shows. Indeed, Figure 3 can also be seen as a case of 1-arbitrage.

**Lemma 1.** *If a pricing function $p_{\epsilon,\lambda}$ is* arbitrage-free *in dataset $D$, then it is also error-monotone in $D$.*

*Proof.* Suppose that $p_{\epsilon,\lambda}$ is not error-monotone in $D$. This implies that there exist parameters $\delta_1, \delta_2$ such that $\mathbb{E}\left[\epsilon(\hat{h}_\lambda^{\delta_1}(D), D)\right] \leq \mathbb{E}\left[\epsilon(\hat{h}_\lambda^{\delta_2}(D), D)\right]$ and $p_{\epsilon,\lambda}(\delta_1, D) < p_{\epsilon,\lambda}(\delta_2, D)$. It is easy to see that in this case $p_{\epsilon,\lambda}$ exhibits 1-arbitrage, since we can simply pick the function $g$ to be the identity function. In this case, $p_{\epsilon,\lambda}$ cannot be arbitrage-free. □

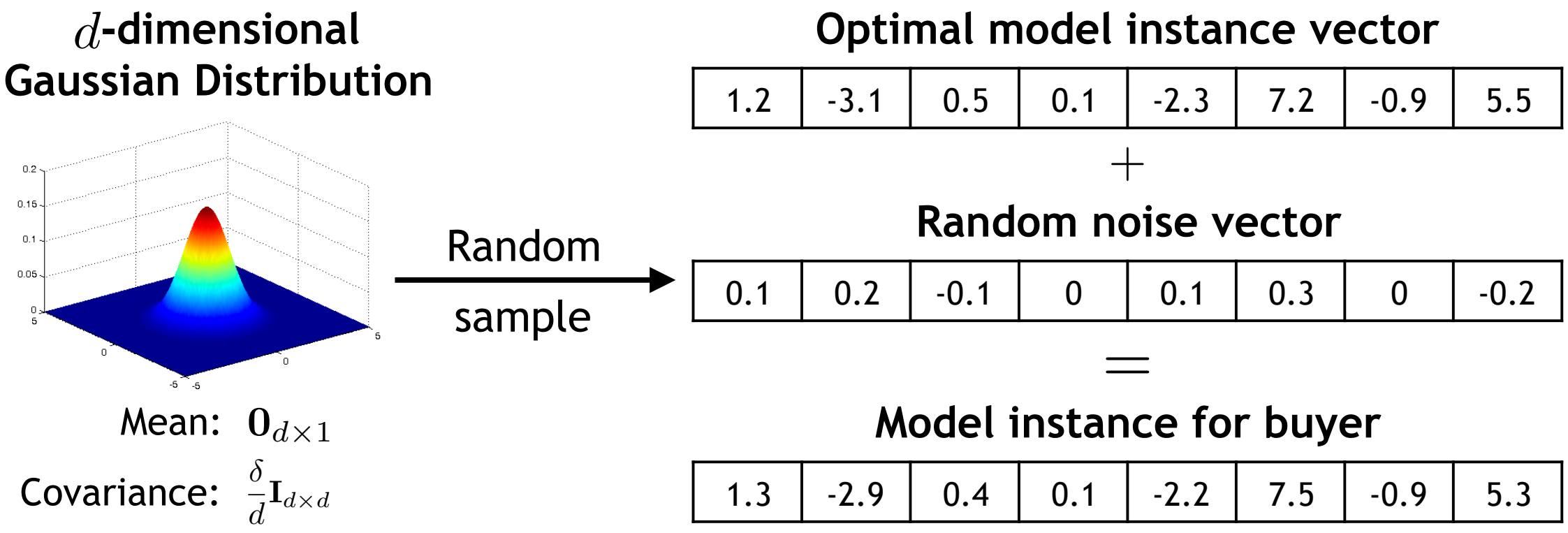

Figure 4: The Gaussian Mechanism for adding random noise to an optimal model instance.

**Definition 5.** *We say that a pricing function $p_{\epsilon,\lambda}$ is* well-behaved *in dataset $D$ iff it is non-negative and arbitrage-free.*

# 4 Noisy Model Generation

So far we have presented a general framework for pricing ML models, and the interactions between the three agents. In this section, we describe a concrete instance of this framework, and present specific mechanisms for noise addition and price computation.

## 4.1 The Gaussian Mechanism

Let us fix a hypothesis space $\mathcal{H}$, such that model instances in $\mathcal{H}$ are vectors in $\mathbb{R}^d$.[1] We will focus on a specific randomized mechanism, denoted $\mathcal{K}_G$, which uses additive Gaussian noise. In particular, define $\mathcal{W}_\delta = \mathcal{N}(\mathbf{0}, (\delta/d) \cdot \mathbf{I}_d)$, for any $\delta \in \mathbb{R}_+$. Here $\mathbf{0}$ is the $d$-dimensional vector with all 0 entries, and $\mathbf{I}_d$ is the identity matrix with dimensions $d \times d$.
Given the two error functions $\epsilon$, $\lambda$, a dataset $D$ and a parameter $\delta$, the Gaussian mechanism first computes the optimal model $h^*_\lambda(D)$ for the given error function $\lambda$ and dataset $D$, samples a vector $w \sim \mathcal{W}_\delta$, and finally outputs $h^*_\lambda(D) + w$. This is illustrated in Figure 4. Formally:

$$\mathcal{K}_G(h^*_\lambda(D), w) = h^*_\lambda(D) + w, \quad w \sim \mathcal{N}(\mathbf{0}, (\delta/d) \cdot \mathbf{I}_d) \tag{1}$$

It is straightforward to see that by construction $\mathcal{K}_G$ is an unbiased mechanism.

**Lemma 2.** *$\mathcal{K}_G$ is an unbiased randomized mechanism.*

*Proof.* Indeed, we have

$$\begin{aligned}\mathbb{E}\left[\mathcal{K}_G(h^*_\lambda(D), w)\right] &= \mathbb{E}\left[h^*_\lambda(D) + w\right] \\ &= h^*_\lambda(D) + \mathbb{E}\left[w\right] = h^*_\lambda(D)\end{aligned}$$

where the last equality comes from the fact that the Gaussian noise we add has mean 0 in every dimension. □

[1] For simplicity, we assume that the dimension of the models equals to the number of features $d$, but note that our framework works in general even if they are not equal.

We next analyze the Gaussian mechanism $\mathcal{K}_G$ in detail. We first focus on a particular instantiation of the error function $\epsilon$, the *square loss*. The square loss computes the error as the euclidean distance from the optimal model:

$$\epsilon_s(h, D) = \|h - h^*_\lambda(D)\|_2^2 \tag{2}$$

When the error function is the square loss, we can show that the parameter $\delta$ is exactly equal to the expected error of the mechanism (and hence trivially the parameter $\delta$ behaves monotonically w.r.t. the expected error).

**Lemma 3.** *Let $\lambda, D$, and $\hat{h}^\delta_\lambda(D) = \mathcal{K}_G(h^*_\lambda(D), w)$. Then:*

$$\mathbb{E}\left[\epsilon_s\left(\hat{h}^\delta_\lambda(D), D\right)\right] = \delta$$

*Proof.* We have:

$$\begin{aligned}\mathbb{E}\left[\epsilon_s\left(\hat{h}^\delta_\lambda(D), D\right)\right] &= \mathbb{E}\left\|\hat{h}^\delta_\lambda(D) - h^*_\lambda(D)\right\|_2^2 \\ &= \mathbb{E}\left[\|w\|_2^2\right] = \sum_{i=1}^{p} \mathbb{E}\left[w_i^2\right] = \delta\end{aligned}$$

This concludes the proof. □

We can show that other types of error functions behave monotonically w.r.t. the expected error.

**Theorem 4.** *Let $D, \lambda$, and let $\epsilon$ be convex as a function of the model instance $h$. Let $\hat{h}^\delta_\lambda(D) = \mathcal{K}_G(h^*_\lambda(D), w)$. Then, for any two parameters $\delta_1, \delta_2$, we have*

$$\mathbb{E}\left[\epsilon(\hat{h}^{\delta_1}_\lambda(D), D)\right] \geq \mathbb{E}\left[\epsilon(\hat{h}^{\delta_2}_\lambda(D), D)\right]$$

*if and only if $\delta_1 \geq \delta_2$.*
*If $\epsilon$ is additionally strictly convex, the above holds with strict inequality ($>$).*

## 4.2 Arbitrage for the Gaussian Mechanism

We now turn our attention to the pricing function that corresponds to the Gaussian mechanism. We show the central theoretical result of this paper, which gives us a concise characterization of an arbitrage-free pricing function when we use the Gaussian mechanism.

**Theorem 5.** *Let $D$ be a dataset, and $\lambda$ be an error function. A pricing function $p_{\epsilon_s,\lambda}$ is arbitrage free for the Gaussian mechanism $\mathcal{K}_G$ if and only if the following two conditions hold for every $\delta_1, \delta_2, \delta_3$:*

1. *If $1/\delta_1 = 1/\delta_2 + 1/\delta_3$, then*

$$p_{\epsilon_s,\lambda}(\delta_1, D) \leq p_{\epsilon_s,\lambda}(\delta_2, D) + p_{\epsilon_s,\lambda}(\delta_3, D).$$

2. *If $\delta_1 \leq \delta_2$, then*

$$p_{\epsilon_s,\lambda}(\delta_1, D) \geq p_{\epsilon_s,\lambda}(\delta_2, D).$$

The above theorem tells us that arbitrage-freeness is equivalent to the function $\bar{p}(x) = p_{\epsilon_s,\lambda}(1/x, D)$ being *subadditive* and *monotone* over its domain. Hence, we have a concise criterion to check for the arbitrage freeness property in a given pricing function $p$.
Although the square loss gives us a compact theoretical characterization of arbitrage-freeness, it is not typically used to measure the error of the model instance returned to the user. However, in the case where $\epsilon$ is a strictly convex function, we can still characterize arbitrage-freeness by applying Theorem 4. Indeed, as a corollary of Theorem 4, if $\epsilon$ is strictly convex, there exists a bijection between the expected error and the parameter $\delta$. Thus, there exists a function $\phi$, which we call the *error-inverse* of $\epsilon$, such that:

$$\delta = \phi\left(\mathbb{E}\left[\epsilon(\hat{h}^{\delta}_{\lambda}(D), D)\right]\right)$$

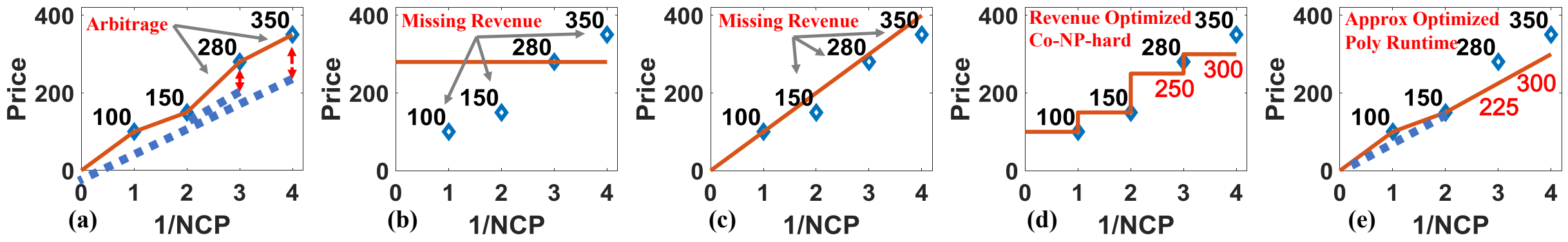

Figure 5: Illustrating example of revenue optimization. Consider a revenue maximization problem with 4 points. $a_1 = 1, a_2 = 2, a_3 = 3, a_4 = 4, b_1 = b_2 = b_3 = b_4 = 0.25, v_1 = 100, v_2 = 150, v_3 = 280, v_4 = 350$. (a) sets all prices equal to the valuation, but it has arbitrage issue. (b) and (c) use constant and linear pricing functions, respectively. They avoid arbitrage, but they lose revenue. (d) gives the revenue-optimal pricing function, which is coNP-hard to compute. (e) is the proposed pricing function that approximates the optimal revenue well while it can be efficiently computed.

We can combine the above insight with Theorem 5 to show the following result.

**Theorem 6.** *Let $D$ be a dataset, and $\lambda, \epsilon$ be error functions. Suppose that $\epsilon$ is strictly convex, and let $\phi$ be its error-inverse. A pricing function $p_{\epsilon,\lambda}$ is arbitrage free for the Gaussian mechanism $\mathcal{K}_G$ if and only if the function*

$$\bar{p}(x) = p_{\epsilon,\lambda}(1/\phi(x), D)$$

*is monotone and subadditive.*

In other words, arbitrage-freeness is still characterized through monotonicity and subadditivity once we view the pricing function through the transformation of the inverse map $\phi$. A natural question here is how one can compute the error inverse $\phi$ of a given $\epsilon$. In general, we can always compute $\phi$ empirically, but in several cases it is possible to compute it analytically.

# 5 Revenue Optimization

So far we have introduced the gaussian mechanism to offer for sale noisy ML models to the buyer, and showed a simple characterization of when a pricing function is arbitrage-free under this mechanism. In this section, we will study the question of how we can assign arbitrage-free prices, with the goal of maximizing the seller's revenue.
Throughout this section, we fix a dataset $D$, the two error functions $\epsilon$ (which is strictly convex) and $\lambda$, and consider only the gaussian mechanism $\mathcal{K}_G$. Instead of dealing directly with the pricing function $p_{\epsilon,\lambda}$, it will be more convenient to express the price as $\hat{p}(x) = p_{\epsilon,\lambda}(1/\phi(x), D)$, where $\phi$ is the error-inverse of $\epsilon$.
Recall that the MBP framework sells models of different *versions*, where each version is parametrized by the NCP $x$ that controls the error. We next describe two specific scenarios of price setting, and then provide a general formalism that captures both.

***Price Interpolation.*** Suppose that the seller wants to set the pricing function such that it takes specific prices for a set of parameters. In particular, the seller provides $n$ *price points* of the form $(a_j, P_j)$, where $a_j$ is a parameter value, and $P_j$ its desired price. The goal is to find an arbitrage-free and non-negative pricing function such that the values $\hat{p}(a_j)$ are as close as possible to $P_j$.

We can capture the above setting by solving the problem of finding an arbitrage-free and non-negative pricing function that maximizes the following objective:

$$T_{\text{PI}}(x_1, \ldots, x_n) = -\sum_{j=1}^{n} \ell(x_j, P_j)$$

where $x_j = \hat{p}(a_j)$. Here, $\ell(x, y)$ can be any loss function such that $\ell(x, y) \geq 0$ and $\ell(x, y) = 0$ if and only if $x = y$. For example, we can choose $\ell(x, y) = |x - y|$, or also $\ell(x, y) = (x - y)^2$, in which case we obtain the functions $T_{\text{PI}}^{\infty}$ and $T_{\text{PI}}^{2}$ respectively.

***Revenue Maximization from Buyer Valuations.*** Assume that the buyers who are interested in buying a model with parameter $x$ have a *valuation* $v_x$ for this model. This implies that they will buy the model only if $\hat{p}(x) \leq v_x$. Moreover, we can capture "how many" buyers are interested in the particular model with parameter $a_j$ through a parameter $b_j$. In this setting, the profit of the seller setting the price at $\hat{p}(a_j) = x$ is $b_j x \cdot \mathbf{1}_{x \leq v_j}$, where $\mathbf{1}_{x \leq v_j}$ is an indicator variable that takes value 1 if $x \leq v_j$, otherwise 0.
Suppose that the seller through market research has obtained the values $v_j, b_j$ for $n$ of these parameters (which correspond to the demand and value curves in Figure 2(a)). We can capture this setting by solving the problem of finding an arbitrage-free and non-negative pricing function that maximizes the following objective:

$$T_{\text{BV}}(x_1, \ldots, x_n) = \sum_{j=1}^{n} b_j x_j \cdot \mathbf{1}_{x_j \leq v_j}$$

where again $x_j = \hat{p}(a_j)$.

We can capture both scenarios by a general optimization problem. Specifically, we are given $n$ *parameter points* $\{a_1, \ldots, a_n\}$, and an objective function $T$. The goal is to find a function $\hat{p}$ that maximizes the quantity $T(\hat{p}(a_1), \ldots, \hat{p}(a_n))$ such that $\hat{p}$ is a well-behaved pricing function, *i.e.*, it is arbitrage-free and non-negative. Formally, we want to solve the following optimization problem:

$$\begin{aligned} \mathbf{max}_{\hat{p}} \quad & T(\hat{p}(a_1), \ldots, \hat{p}(a_n)) \\ \mathbf{subject\ to} \quad & \hat{p}(x + y) \leq \hat{p}(x) + \hat{p}(y), \qquad x, y \geq 0 \\ & \hat{p}(y) \geq \hat{p}(x), \qquad y \geq x \geq 0 \\ & \hat{p}(x) \geq 0, \qquad x \geq 0 \end{aligned} \tag{3}$$

The first two constraints in (3) capture the subadditivity and monotonicity constraints that result from the arbitrage-freeness requirement. The third constraint corresponds to the non-negative requirement. Observe that the above optimization problem is not over a set of variables, but over the space of all functions $\hat{p}$.

## 5.1 Hardness Results

We now study the computational complexity of solving the optimization problem (3). We will show that the problem is intractable for all objective functions $T_{\text{BV}}, T_{\text{PI}}^{\infty}, T_{\text{PI}}^{2}$ that we defined in the previous section. To show this hardness result, we first consider a decision problem that we call SUBADDITIVE INTERPOLATION.

**Definition 6** (SUBADDITIVE INTERPOLATION). *Given as input a set of points $\{(a_j, P_j)\}_{j=1}^{n}$, where $a_j, P_j$ are non-negative rational numbers, does there exists a function $\hat{p}$ that $(i)$ is positive, monotone and subadditive, and $(ii)$ satisfies $\hat{p}(a_j) = P_j$.*

We show in the Appendix that SUBADDITIVE INTERPOLATION is a computationally hard problem.

**Theorem 7.** SUBADDITIVE INTERPOLATION *is co$\mathcal{NP}$-hard.*

Equipped with Theorem 7, we can show that the other optimization problems are also hard. Indeed, suppose that the objective function $T$ has a unique maximizer, $(\theta_1, \dots, \theta_n)$. Then, the optimization problem (3) will return the maximum value of $T$ if and only if for every $j = 1, \dots, n$ we have $\hat{p}(a_j) = \theta_j$. We can use this observation to prove the following result:

**Corollary 7.1.** *The optimization problem* (3) *with objective functions any of* $\{T_{\text{BV}}, T_{\text{PI}}^{\infty}, T_{\text{PI}}^{2}\}$ *is co*$\mathcal{NP}$*-hard.*

*Proof.* Observe that the objective functions $T_{\text{PI}}^{\infty}, T_{\text{PI}}^{2}$ are maximized if and only if $x_j = P_j$ for $j = 1, \dots, n$. Hence, we can reduce SUBADDITIVE INTERPOLATION to (3). Similarly, the objective functions $T_{\text{BV}}$ is maximized at the unique point where $x_j = v_j$, *i.e.* the price at $a_j$ equals the valuation $v_j$. Hence, we can reduce SUBADDITIVE INTERPOLATION to the revenue maximization with buyer valuations problem by setting $v_j = P_j$ and $b_j = 1$ for every $j = 1, \dots, n$. □

## 5.2 Approximating Subadditivity

In order to overcome the hardness of the original optimization problem (3), we seek to approximately solve it by modifying the subadditivity constraint. In particular, we replace the subadditive constraints $\hat{p}(x+y) \leq \hat{p}(x) + \hat{p}(y)$ by the constraints $\hat{q}(x)/x \geq \hat{q}(y)/y$ for every $0 < x \leq y$. In other words, we want to find $\hat{q}$ such that $\hat{q}(x)/x$ is a decreasing function of $x$. Geometrically, one can think of this condition as that the slope of the line that connects the origin with the point $(x, q(x))$ is non-increasing. The reformulated optimization problem is as follows:

$$
\begin{array}{rll}
\textbf{max}_{\hat{q}} & T(\hat{q}(a_1), \dots, \hat{q}(a_n)) & \\
\textbf{subject to} & \hat{q}(y)/y \leq \hat{q}(x)/x, \quad y \geq x > 0 & \\
& \hat{q}(y) \geq \hat{q}(x), \quad y \geq x \geq 0 & \\
& \hat{q}(x) \geq 0, \quad x \geq 0 &
\end{array}
\tag{4}
$$

It is easy to show that for any feasible solution $\hat{q}$ of (4), the pricing function $\hat{p}(x) = \hat{q}(x)$ is also a feasible solution of (3) and hence $\hat{q}$ is a well-behaved pricing function as well.

**Lemma 8.** *Any pricing function $\hat{q}$ that satisfies the constraints of* (4) *is arbitrage-free and non-negative.*

***Approximation Guarantees.*** We can show that for any pricing function $\hat{p}$ that is a feasible solution of (3), we can find a feasible solution $\hat{q}$ of (4) that is not too far away from $\hat{p}$. We will use this fact later to show that our approximation does not lose too much from the optimal objective value. More precisely:

**Lemma 9.** *Let $\hat{p}$ be a feasible solution of* (3)*. Then, there exists a feasible solution $\hat{q}$ of* (4) *such that for every $x > 0$:*

$$\hat{p}(x)/2 \leq \hat{q}(x) \leq \hat{p}(x)$$

***From Functions to Variables.*** The resulting optimization problem in (4) is still over the space of all possible functions. However, it turns out that we can equivalently rewrite it so that it searches over variables instead of functions. The key observation is that we only need to find the values of the function $\hat{q}$ only for the $n$ parameter points $a_1, \dots, a_n$. In particular, consider the following optimization problem.

$$
\begin{array}{rll}
\textbf{max}_{\mathbf{z}} & T(z_1, \dots, z_n) & \\
\textbf{subject to} & z_j/a_j \leq z_i/a_i, \quad a_j \geq a_i & \\
& z_j \geq z_i, \quad a_j \geq a_i & \\
& z_j \geq 0, \quad 1 \leq j \leq n &
\end{array}
\tag{5}
$$

The next proposition tells us that the two formulations are essentially equivalent. In particular, if $\mathbf{z}^*$ is an optimal solution for (5), then we can use it to construct an optimal solution $\hat{q}$ for (4). The construction is simple: we define $\hat{q}$ as the piecewise linear function that goes through the points $(a_i, z_i)$.

**Proposition 1.** *For every feasible solution of* (4)*, there exists a feasible solution of* (5) *with the same value of the objective function, and vice versa.*

We conclude this section by providing an example of the construction of the approximate optimization program.

**Example 3.** *Consider the revenue maximization problem with parameters as given in Figure 5. In this scenario, the optimization problem 5 can be written as follows:*

$$\begin{aligned} &\textbf{max} \quad T_{\text{BV}}(z_1, z_2, z_3, z_4) \\ &\textbf{subject to} \quad z_1 \geq z_2/2 \geq z_3/3 \geq z_4/4 \\ &\qquad z_4 \geq z_3 \geq z_2 \geq z_1 \geq 0 \end{aligned}$$

## 5.3 Algorithms for Revenue Optimization

In this section, we show how to leverage the approximate optimization problem (5) in order to obtain efficient algorithms with some approximation guarantees. The advantage of (5) is that it is an optimization problem with linear constraints. Depending on the objective function $T$, this problem can be tractable. For example, if $T$ is concave, then (5) is a linear program with a concave objective function, which can be solved in polynomial time.

We will focus on the two problems we introduced before: price interpolation, and revenue maximization from buyer valuations.

***Price Interpolation.*** It is easy to see that both objective functions $T_{\text{PI}}^{\infty}, T_{\text{PI}}^{2}$ are concave. This implies immediately that (5) with the above two objective functions can be solved in polynomial time.

We next show that we can also obtain an (additive) approximation guarantee. For this, we need the following general result:

**Proposition 2.** *Let $C_{MPB}, C_{SA}$ be the optimal values of* (5) *and* (3) *respectively under an objective function $T(z_1, \ldots, z_n) = \sum_{i=1}^{n} T_i(z_i)$, where each $T_i$ is concave and non-positive. Then,*

$$C_{SA} + \sum_i T_i(0)/2 \leq C_{MBP} \leq C_{SA}$$

We can apply the above proposition for the objective function $T_{\text{PI}}^{\infty}$: this implies that the optimal value of the approximate solution will be at most $(\sum_j P_j)/2$ away from the optimal solution. Similarly, we can obtain a additive approximation guarantee of $(\sum_j P_j^2)/2$ for $T_{\text{PI}}^{2}$.

***Revenue Maximization from Buyer Valuations.*** In contrast to price interpolation, $T_{\text{BV}}$ is not a concave function. However, we will show that (5) can still be solved in polynomial time, and moreover, that the optimal solution is within a constant approximation factor of the optimal solution of (3).

We first show that approximating the subadditive constraints loses at most a factor of $1/2$ when the objective function is $T_{\text{BV}}$.

**Proposition 3.** *Let $C_{MPB}, C_{SA}$ be the optimal values of* (5) *and* (3) *respectively under the objective function $T_{\text{BV}}$. Then,*

$$C_{SA}/2 \leq C_{MBP} \leq C_{SA}$$

Next, we provide an algorithm based on dynamic programming that optimally solves (5) with objective function $T_{\text{BV}}$.

Suppose that we are given as input the parameters $v_j, b_j$ that correspond to point $a_j$ for $j = 1, \ldots, n$. We assume that $a_1 \leq a_2 \leq \cdots \leq a_n$ and $v_1 \leq v_2 \leq \cdots \leq v_n$ (*i.e.*, the valuations of the buyers are monotone w.r.t. the error).

Let $s(k, \Delta)$ denote the optimum solution for the subproblem with points $j = k, \dots, n$, with the restriction that for every $j \geq k$ we have $s_j(k, \Delta)/a_j \leq \Delta$. We will denote by $OPT(k, \Delta)$ the objective value of this optimum solution. Observe that the optimum solution for the initial problem is simply $s(1, +\infty)$, with optimum value $OPT(1, +\infty)$.
We will provide a recursive formula to compute $OPT(k, \Delta)$ for any $k, \Delta$. Our first observation is that for $k = n$, we can easily compute the optimum solution as follows:

$$s_n(n, \Delta) = \min\{v_n, \Delta a_n\}$$
$$OPT(n, \Delta) = b_n \cdot s_n(n, \Delta)$$

This follows from the fact that it is always more profitable to assign a higher price, as long as it is under the valuation $v_n$.
We also need the following lemma.

**Lemma 10.** *For every $k$, $s_k(k, \Delta) \geq \min\{v_k, \Delta a_k\}$.*

*Proof.* Suppose not; we will then show that we can obtain a solution with a larger objective value. Let $\ell \geq k$ be the largest index such that $s_\ell(k, \Delta) = s_k(k, \Delta)$. Clearly we have that $s_k(k, \Delta) = s_{k+1}(k, \Delta) = \dots, s_\ell(k, \Delta)$. Since $s_k(k, \Delta) < v_k$ and $v_k \leq v_{k+1} \leq \dots \leq v_\ell$, we must have that $s_j(k, \Delta) < v_j$ for every $j = k, \dots, \ell$. Similarly, since $s_k(k, \Delta) < \Delta a_k$ and $a_k \leq a_{k+1} \leq \dots \leq a_\ell$, we must have that $s_j(k, \Delta) < \Delta a_j$ for every $j = k, \dots, \ell$.
Let $\epsilon = \min_{j=k}^{\ell} \frac{\min\{v_k, \Delta a_k\}}{s_j(k,\Delta)} > 1$. Define $s'(k, \Delta)$ such that for $j = k, \dots, \ell$ we have $s'_j(k, \Delta) = \epsilon \cdot s(k, \Delta)$, and for $j > \ell$ it remains the same. It is easy to see that the resulting solution is feasible, and also produces a strictly greater revenue, a contradiction. □

To compute the recursive formula for $s(k, \Delta)$, we distinguish between two cases.

**Lemma 11.** *Let $k < n$. If $a_k \Delta \leq v_k$, then:*

$$s_k(k, \Delta) = \Delta a_k,\ s_j(k, \Delta) = s_j(k + 1, \Delta), j > k$$
$$OPT(k, \Delta) = b_k \Delta a_k + OPT(k + 1, \Delta)$$

*Proof.* From Lemma 10, we have that $s_k(k, \Delta) \geq \Delta a_k$. But it must also be that $s_k(k, \Delta) \leq \Delta a_k$, thus the only optimal solution is $s_k(k, \Delta) = \Delta a_k$.
Moreover, since $(\Delta a_k)/a_k = \Delta \geq s_j(k + 1, \Delta)/a_{k+1}$, the weakened subadditive constraint is satisfied. Finally, we can write $s_{k+1}(k + 1, \Delta) \geq \min\{\Delta a_{k+1}, v_{k+1}\} \geq \min\{\Delta a_k, v_k\} = \Delta a_k$, which means that monotonicity is also satisfied. □

**Lemma 12.** *Let $k < n$. If $a_k \Delta > v_k$, define*

$$s'_k(k, \Delta) = v_k,\ s'_j(k, \Delta) = s_j(k + 1, v_k/a_k), j > k$$
$$s''_k(k, \Delta) = s_{k+1}(k + 1, \Delta)\frac{a_k}{a_{k+1}},\ s''_j(k, \Delta) = s_j(k + 1, \Delta), j > k$$

*with optimum values respectively*

$$OPT'(k, \Delta) = b_k v_k + OPT(k + 1, v_k/a_k)$$
$$OPT''(k, \Delta) = OPT(k + 1, \Delta)$$

*Then, $OPT(k, \Delta)$ is the maximum between the two options, and $s(k, \Delta)$ is the solution that achieves the maximum.*

Table 3: Dataset Statistics.

| Task | DataSet | $n_1$ | $n_2$ | $d$ |
|---|---|---|---|---|
| Regression | Simulated1 | 7500000 | 2500000 | 20 |
| | YearMSD | 386509 | 128836 | 90 |
| | CASP | 34298 | 11433 | 9 |
| Classification | Simulated2 | 7500000 | 2500000 | 20 |
| | CovType | 435759 | 145253 | 54 |
| | SUSY | 3750000 | 1250000 | 18 |

*Proof.* From Lemma 10, we have that $s_k(k, \Delta) \geq v_k$. The first option examines what will happen if we set $s_k(k, \Delta) = v_k$, in which case we will obtain a profit of $b_k v_k$ from this price point. If $s_k(k, \Delta) > v_k$, then we obtain 0 revenue from this point. Also, the more we increase the price, the more revenue we can extract from the remaining price points until we reach $\Delta a_k$.
It is straightforward to see that the weakened subadditive constraint is satisfied in both cases. We now show the same for monotonicity as well.
For the first option, since $v_k \leq v_{k+1}$ and $v_k = (v_k/a_k)a_k \leq (v_k/a_k)a_{k+1}$, we can write:

$$v_k \leq \min\{v_{k+1}, (v_k/a_k)a_{k+1}\} \leq s_{k+1}(k+1, v_k/a_k)$$

For the second option, monotonicity follows from the fact that $a_k \leq a_{k+1}$ □

We can now use the recursive formulas from the two lemmas to obtain an efficient dynamic programming algorithm. The key observation is that we only need to consider $(n+1)$ values of $\Delta$, since from the recurrence relations, $\Delta$ can only take values from the set $\{v_1/a_1, v_2/a_2, \ldots, v_n/a_n, +\infty\}$. The dynamic programming algorithm will first compute $s(n, \Delta)$, for the $(n+1)$ values of $\Delta$, and then iteratively compute $s(k, \Delta)$ for $k = n-1, n-2, \ldots, 1$. The final solution will be $s(1, +\infty)$. Since we have $n$ iterations, where each iteration computes $(n+1)$ subproblems, the running time of the algorithm is $O(n^2)$.

**Theorem 13.** *There exists a dynamic programming algorithm that computes the optimal values of* (5) *under the objective function* $T_{\text{BV}}$ *in time* $O(n^2)$.

# 6 Experiments

Our goal of this experimental section is three-fold: (1) validate that the ML model accuracy/error is monotone with respect to the inverse of the noise control parameter 1/NCP (which is the variance of the gaussian noise), (2) show that the MBP framework provides sellers with more revenue while more buyers have access to ML models, and (3) justify that MBP runs much faster than a naive brute-force search for the revenue optimization problem, while still providing near-optimal revenue.
**Experimental Setup**. All experiments were run on a machine with 4 Intel i5-6600 3.3 GHz cores, 16 GB RAM, and 500 GB disk with Ubuntu 14.04 LTS as the OS. We have prototyped the model-based pricing framework in Matlab 2017b.

## 6.1 Expected Error to 1/NCP Transformation

The first natural question is whether it is true that the the expected model accuracy/error is always monotone as a function of 1/NCP, *i.e.*, the inverse of the noise control parameter? Theorem 4 in Section 4 provably gives a positive answer when the error function is strictly convex. This section provides an empirical study on how the expected ML model error varies w.r.t. to the noise control parameter.
We use six datasets that are summarized in Table 3. The first three datasets, Simulated1, YearMSD, and CASP are for the regression task (Linear Regression), while the next three datasets, Simulated2, CovType and SUSY are for the classification task (Logistic Regression). The feature vectors of the dataset Simulated1 and Simulated2 are generated from a normal distribution. The target values of Simulated1 are simply the inner product of the feature vectors and a hyperplane vector. The label value of a data point from Simulated2 is 1 with probability 0.95 if it is above a given hyperplane, and 0 with probability if it is below the hyperplane. The other datasets are all from the UCI machine learning repository [7]. For each value of the NCP, we generate 2000 random models, each of which is equal to the optimal model plus an independently randomly generated vector with the same variance.

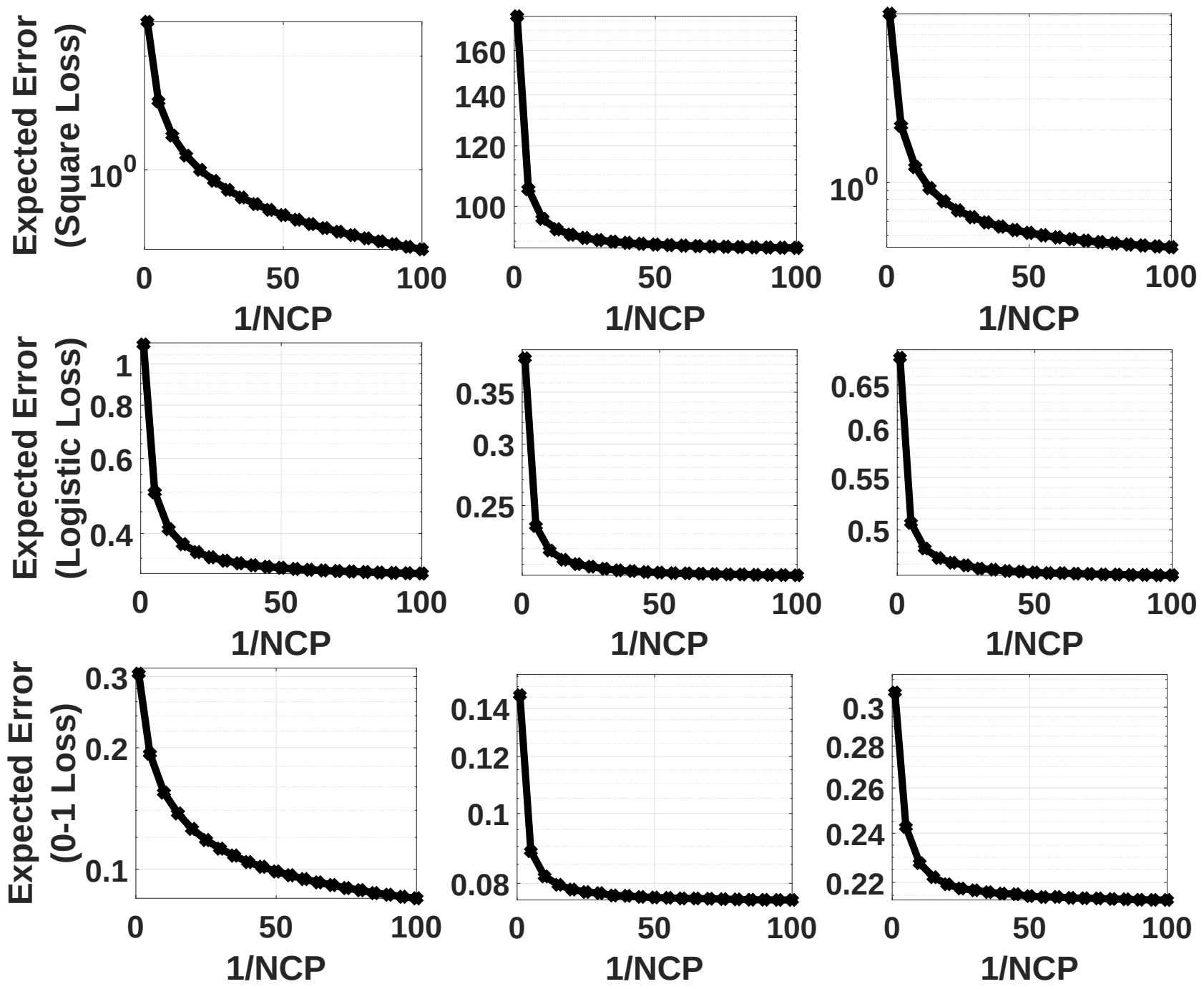

Figure 6: Error Transformation Curve. All errors are measured on the testing datasets. The first row corresponds to the square loss for Simulated1, YearMSD, and CASP, respectively. The second row represents the logistic loss, while the third row shows the 0/1 classification erro for Simulated2, CovType, and SUSY.

As shown in Figure 6, the testing error decreases as the variance inverse increases. This verifies that there is a monotone mapping and hence the error transformation is feasible. Interestingly, even when the error is not strictly convex, such as the 0/1 classification error, the expected error still decreases as 1/NCP increases. This might be because all model instances are trained and tested on a relatively large datasets and thus have good generalization error. Thus, the strictly convex loss function can exactly indicate the 0/1 classification error. In other words, they are monotone to each other and the 0/1 classification error is also monotone to 1/NCP. Note that as 1/NCP increases, the error function first drops sharply and then decreases slowly. This is because improving the error performance is becoming harder and harder as the model instance is closer to the optimal model. For example, it is almost always much more difficult to increase the classification accuracy from 90% to 95% compared to that from 60% to 65%.
In the following part of the experiments, we will focus on 1/NCP, which is monotone and thus represents the expected error.

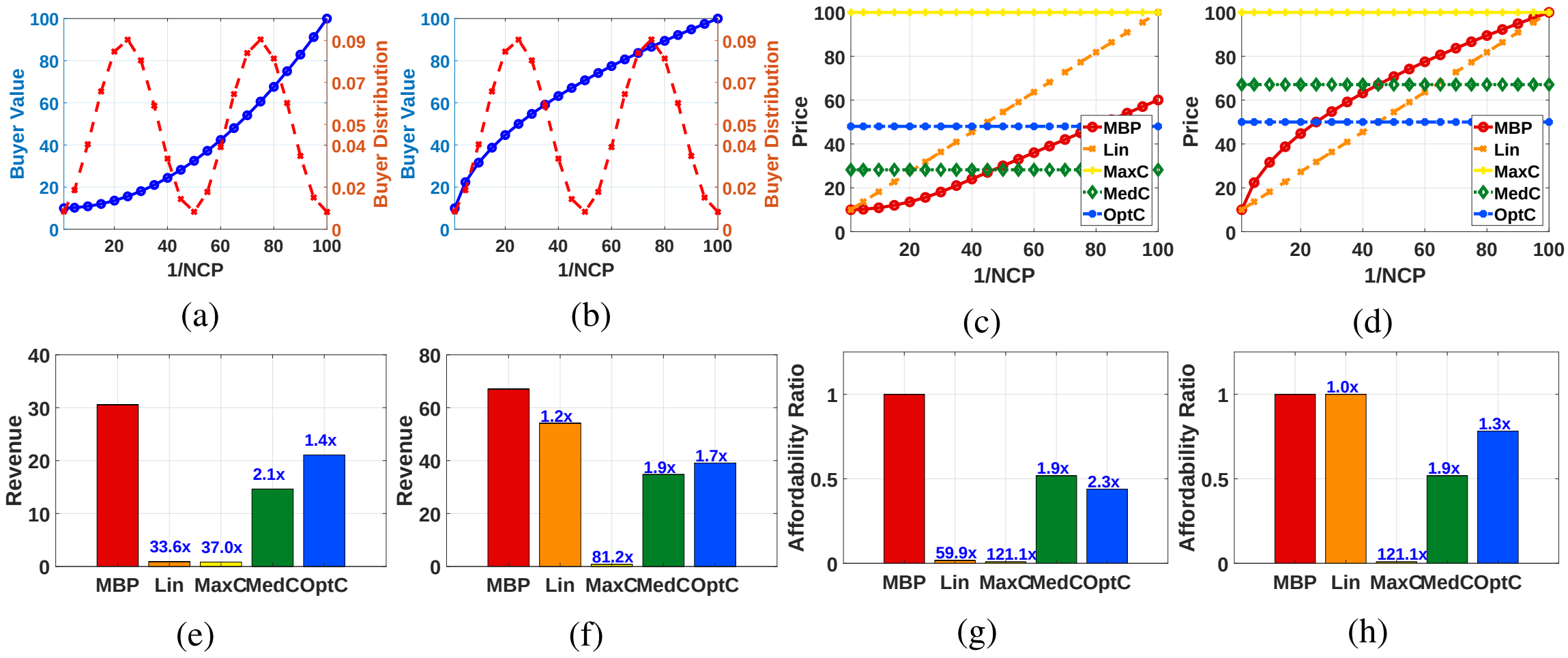

Figure 7: Revenue and Affordability Gain. The buyer distribution is fixed and we vary the buyer value curve.

## 6.2 Revenue and Affordability Gain

Next we study the benefits of our proposed MBP approach on the seller's revenue and buyer's affordability ratio (fraction of the buyers that can afford to buy a model instance) compared to other approaches of pricing ML models. We consider the setting of revenue maximization from buyer valuations as this is described in Section 5, *i.e.*, a buyer would pay for a model instance if and only if the price is less than the buyer's valuations. We compare MBP with four pricing approaches, namely, Lin, MaxC, MedC, and OptC (all of which obtain well-behaved pricing functions). Lin, the linear approach, uses a linear interpolation of the smallest and largest value in the buyer's value curve to set the price. MaxC, MedC, and OptC set a single price for all ML model instances. MaxC uses the highest value in the buyer's value curve. MedC uses a price such that half of the buyer can afford to buy a model instance. OptC uses a constant price which maximizes the seller's profit.

Figure 7 and 8 show the results under different buyer value and demand curves, respectively. Overall, MBP can achieve up to 81.2x revenue gains and up to 121.1x affordability gains compared to the four baseline approaches.

We first fix the buyer distribution and vary the buyer valuation. As shown in Figure 7 (a), when the value curve is convex, MBP obtains significantly more revenue and affordability compared to the linear approach. This is because the linear approach misses the opportunities to sell model instances to buyers interested in buying model instances with medium accuracy. When the buyer curve becomes concave as shown in Figure 7 (b), however, the linear approach can achieve more revenue and affordability as more buyers can be satisfied. Nevertheless, the constant approaches now suffer from losing revenue as they cannot accordingly change the price for different buyers. Meanwhile, MBP achieves the largest revenue gains and affordability, as a concave function is also a subadditive function and thus MBP can match exactly the value curve.

Next, we fix the buyer value curve and vary the demand curve. As shown in Figure 8, when most of the buyers are interested in buying model instances with medium accuracy, MBP tends to produce a price function that ties close to the price for model instance with medium accuracy. When most buyers are interested in buying extremely low and extremely high accurate model instances, MBP can accordingly change the price function it generates to follow the different requirement. Meanwhile, as shown in Figure 8 (c) and (d), none of Lin, MaxC, and MedC is able to capture this. While OPTC does change its price function, the effect it has is limited, as it only produces a single price for all model instances. This is why MBP can always achieves the largest revenue gain and affordability ratio.

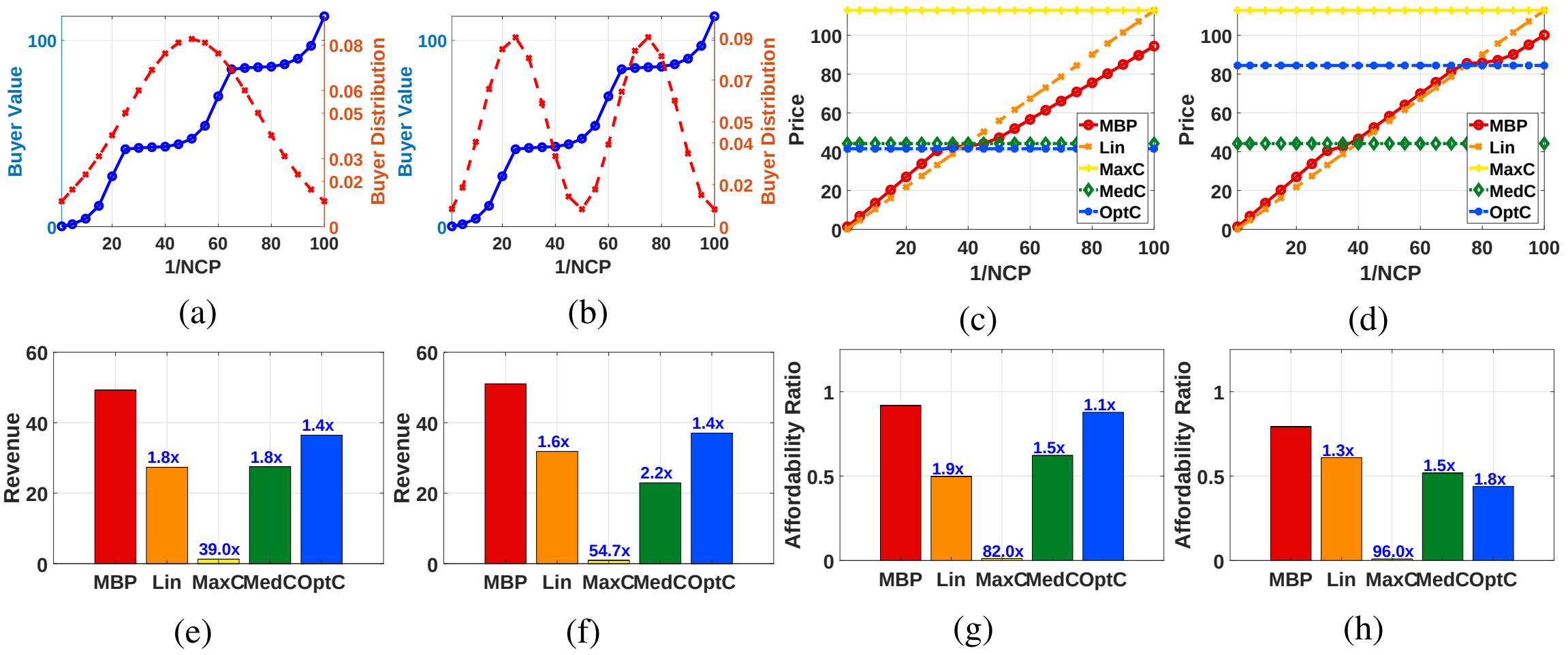

Figure 8: Revenue and Affordability Gain. We fix the buyer valuation and vary the buyer distribution.

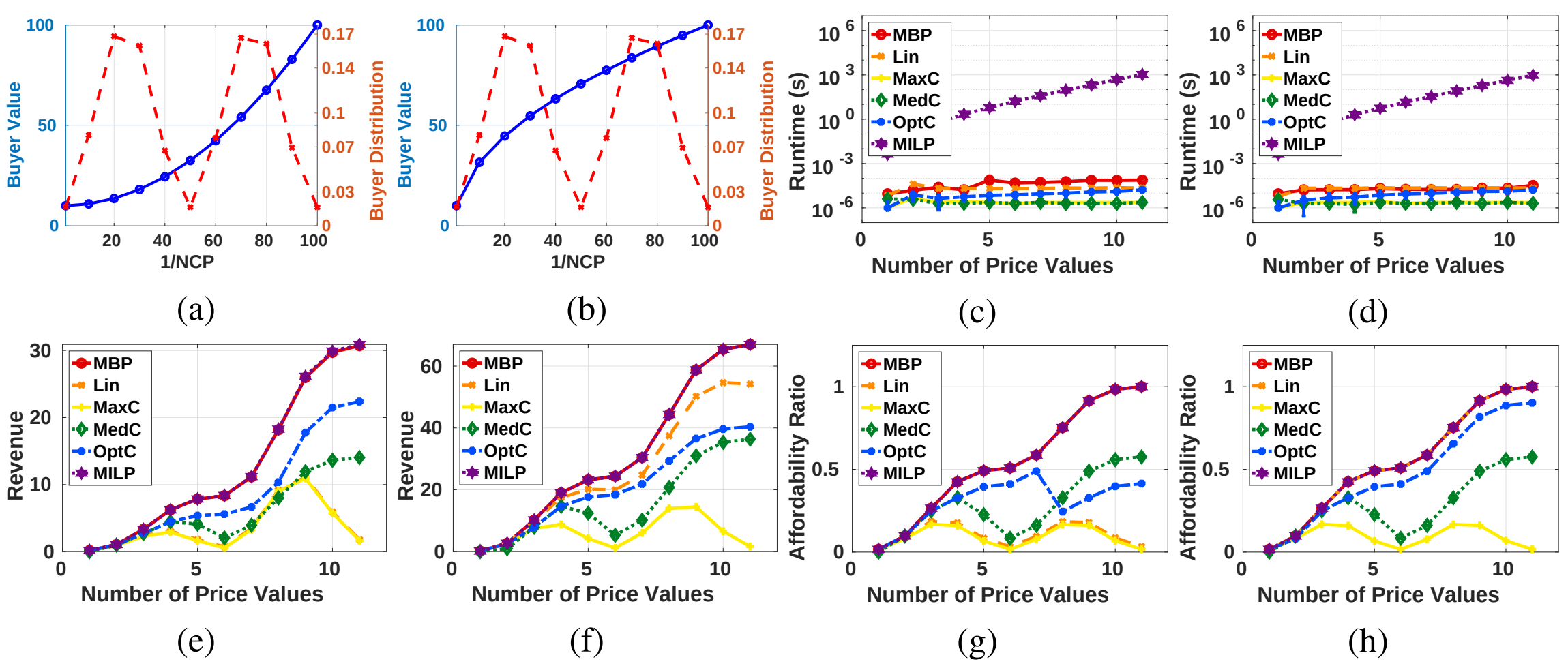

Figure 9: Runtime performance of MBP. We fix the buyer distribution and vary buyer valuation.

### 6.3 Runtime Performance

Finally, we present experimental results on the runtime performance of the revenue optimization algorithms under MBP. Fixing the buyer curve, we vary the number of pricing points and compare the runtime and revenue gains of our MBP proposed, versus the optimal yet expensive optimal algorithm MILP (a multiple-integer-linear programming approach given in the appendix), as well as all the other four baseline methods.

Figures 9 and 10 present how the runtime, revenue, and affordability ratio vary as the buyer distribution and value curve change. Overall, MBP is always more than several orders of magnitude faster than the naive MILP. This is because MILP requires solving integer linear programming exponentially many times. When the number of parameter points/price values increases, the runtime of MILP grows quickly. Since MBP is an algorithm requiring only quadratic runtime, its runtime is much faster. While other naive pricing methods are slightly faster than MBP due to their simplicity, they almost always suffer from either revenue gains or affordability ratio, or both. Note that we have not optimized the implementation of the dynamic revenue optimization algorithm. It is interesting future work to see how much speedup can be obtained by implementing MBP in low level languages such as C/C++ with extensive leverage of CPU, GPU, and IO

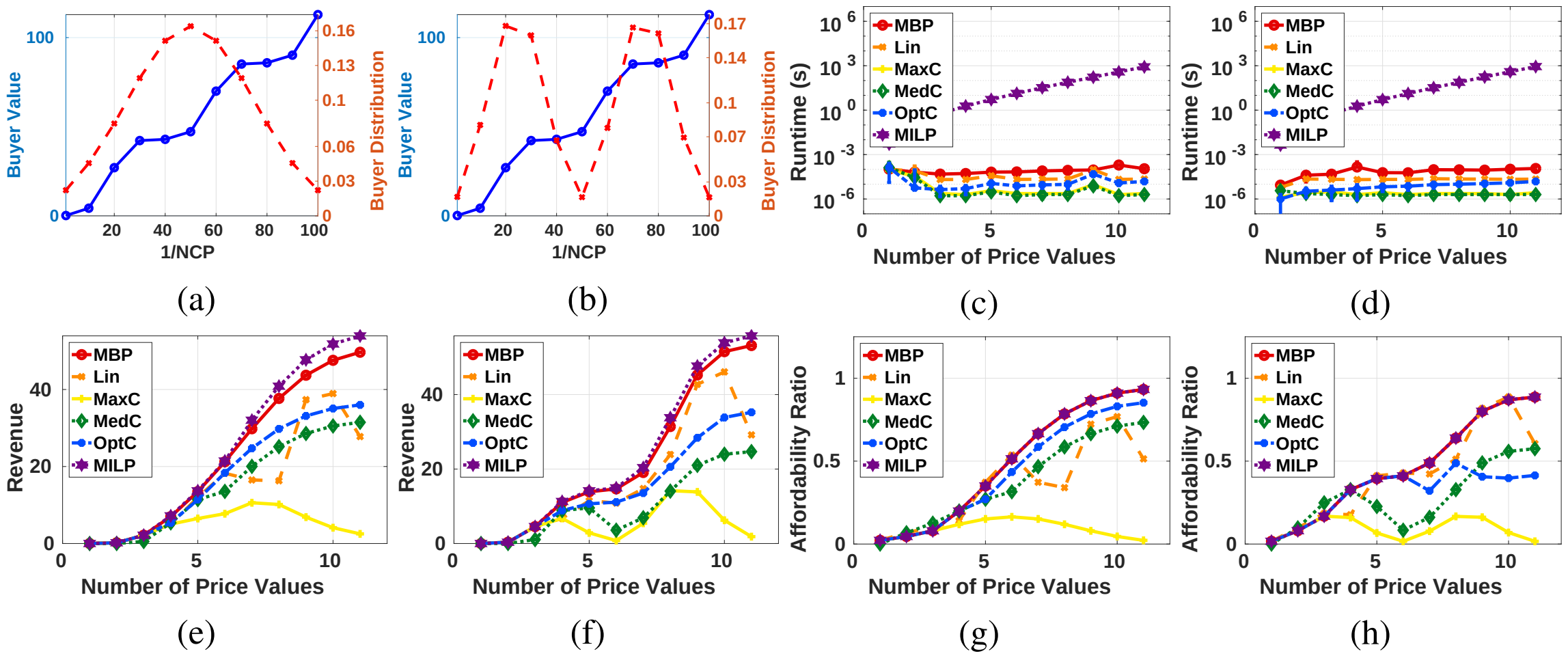

Figure 10: Runtime performance of MBP. We fix buyer value and vary buyer distribution.

optimization.

Furthermore, an interesting phenomenon is that even if MBP theoretically can have a revenue about half of the revenue achieved by MILP, among all experiments we conducted, its revenue is very close to that of MILP, verified by Figure 9 (e) (f) as well as 10 (e) (f). This is because for most of the common buyer valuation curves, the optimal sub-additive curves would de facto be equal or close to, some curves that satisfy the approximated subadditive constraints.

Finally, as shown by Figure 9 (g) (h) and 10 (g) (h), while MILP and MBP do not explicitly optimize the affordability ratio, they almost always produce a pricing curve with highest affordability ratio. This is because, informally speaking, optimizing revenue can be achieved by selling models to as many as possible buyers, which implicitly optimizes the affordability ratio. Nevertheless, in some case, for example, when there are only 3 price values as shown in Figure 10 (h), MedC can achieve an affordability ratio slightly larger than that achieved by MBP and MILP. This is because MedC explicitly requires the affordability ratio larger than 50%. This indicates that there is still room to improve fairness. Due to space limit, we leave a formal study of trade-off between revenue and fairness to future work.

# 7 Conclusion and Future Work

In this work, we initiate the formal study of data markets that sell directly ML models to buyers. We propose a model-based pricing (MBP) framework, which instead of pricing the data, directly prices ML model instances. We show that a concrete realization of the MBP framework via a random noise injection approach provably satisfies several desired formal properties, including preventing arbitrage opportunities. Based on the proposed framework, we then provide algorithmic solutions on how sellers can assign prices to models under different market scenarios (such as to maximize revenue). Extensive experiments validate that the MBP framework can provide high revenue to the sellers, high affordability to the buyers, and can also operate on low runtime cost.

There are several other exciting directions for future work. First, more complex ML models such as Bayesian networks, artificial neural networks, SVMs, and statistical relational models are also frequently used. Non-relational data (images, text, etc.) might require complex feature extraction, possibly implicitly within an ML model (as in deep learning [18]). Handling such complex models is a key avenue to extend our framework. Second, we assumed that buyers know which ML model they want. This is a reasonable starting point because most existing cloud ML platforms assume the buyer picks the ML model. But in practice, users often perform *model selection* and explore different ML models [9, 24] and refine their choices iteratively [16].

Incorporating such manual, iterative, or automated model selection and refinement along with pricing is another key avenue to extend our framework. Third, in many cases the data offered for sale comes with privacy constraints, since it has been extracted from private users. Integrating model-based pricing with data privacy is also a core future challenge. Finally, more complicated buyer models as well as trade-offs between revenue and fairness can be further explored in the revenue optimization.

# A Missing Proofs

*of Theorem 4.* We will use the following property of a strictly convex function $\epsilon$. For every $x, y \in \mathbb{R}^p$ we have

$$\epsilon(x+\sigma_1 y) + \epsilon(x-\sigma_1 y) > \epsilon(x+\sigma_2 y) + \epsilon(x-\sigma_2 y)$$

if and only if $\sigma_1 > \sigma_2$, where $\sigma_1, \sigma_2$ are scalars.
To prove the above property, we apply the definition of strict convexity twice with $t = \frac{1}{2}(1+\frac{\sigma_2}{\sigma_1})$ when $\sigma_1 > \sigma_2$

$$t\epsilon(x+\sigma_1 y) + (1-t)\epsilon(x-\sigma_1 y) > \epsilon(x+\sigma_2 y)$$
$$(1-t)\epsilon(x+\sigma_1 y) + t\epsilon(x-\sigma_1 y) > \epsilon(x-\sigma_2 y)$$

and then sum up the two inequalities. The ”if” part is a result of symmetry.
Let $h^* = h^*_\lambda(D)$, and $w_1, w_2$ the two gaussian noise vectors. We can now compute the expectations as follows:

$$\begin{aligned}\mathbb{E}\left[\epsilon(\hat{h}^{\delta_1}_\lambda(D), D)\right] - \mathbb{E}\left[\epsilon(\hat{h}^{\delta_2}_\lambda(D), D)\right] &= \mathbb{E}\left[\epsilon(h^*+w_1)\right] - \mathbb{E}\left[\epsilon(h^*+w_2)\right] \\ &= \int_{-\infty}^{+\infty} \frac{1}{\sqrt{2\pi}}(\epsilon(h^*+\delta_1 y) - \epsilon(h^*+\delta_2 y))e^{-\frac{1}{2}y^2}dy \\ &= \int_{0}^{+\infty} \frac{1}{\sqrt{2\pi}}(\epsilon(h^*+\delta_1 y) + \epsilon(h^*-\delta_1 y) - \epsilon(h^*+\delta_2 y) - \epsilon(h^*-\delta_2 y))e^{-\frac{1}{2}y^2}dy\end{aligned}$$

The last equality comes from splitting the interval $[-\infty, +\infty]$ to two smaller intervals, $[-\infty, 0]$ and $[0, +\infty]$ and then changing the sign of $y$ in the first term. It is now easy to see that the above quantity is strictly positive if and only if $\delta_1 > \delta_2$. □

*of Theorem 5.* We next prove the two directions of the theorem.

($\Longrightarrow$) Suppose that the pricing function $p_{\epsilon_s,\lambda}$ is arbitrage free. Then, by Lemma 1 the pricing function is also error-monotone in $D$, so condition (2) holds.
To show that condition (1) holds as well, consider parameters $\delta_1, \delta_2, \delta_3$ such that $1/\delta_1 = 1/\delta_2 + 1/\delta_3$ and

$$p_{\epsilon_s,\lambda}(\delta_1, D) > p_{\epsilon_s,\lambda}(\delta_2, D) + p_{\epsilon_s,\lambda}(\delta_3, D).$$

We will show in this case that the pricing function violates $k$-arbitrage for $k = 2$. We define the following function $g$ that combines two models:

$$g(\hat{h}^{\delta_2}_\lambda(D), \hat{h}^{\delta_3}_\lambda(D)) = \frac{\delta_1}{\delta_2}\cdot \hat{h}^{\delta_2}_\lambda(D) + \frac{\delta_1}{\delta_3}\cdot \hat{h}^{\delta_3}_\lambda(D)$$

Now, observe that:

$$\tilde{h} = g(\hat{h}^{\delta_2}_\lambda(D), \hat{h}^{\delta_3}_\lambda(D)) = \frac{\delta_1}{\delta_2}\cdot(h^*_\lambda(D)+w_2) + \frac{\delta_1}{\delta_3}\cdot(h^*_\lambda(D)+w_3) = h^*_\lambda(D) + \frac{\delta_1}{\delta_2}\cdot w_2 + \frac{\delta_1}{\delta_3}\cdot w_3$$

Hence, we can compute the expectation

$$\mathbb{E}\left[\epsilon_s(\tilde{h}, D)\right] = \frac{\delta_1^2}{\delta_2^2}\cdot\mathbb{E}\left[w_2^2\right] + \frac{\delta_1^2}{\delta_3^2}\cdot\mathbb{E}\left[w_3^2\right] = \delta_1^2\left(\frac{1}{\delta_2}+\frac{1}{\delta_3}\right) = \delta_1 = \mathbb{E}\left[\epsilon_s(\hat{h}^{\delta_1}_\lambda(D), D)\right]$$

where the last equality comes from Lemma 3. Hence, the pricing function indeed violates 2-arbitrage, a contradiction.

($\Longleftarrow$) We now show the opposite direction of the theorem, *i.e.*, that conditions (1) and (2) imply arbitrage freeness.
To show this, we will use the Cramér-Rao inequality, which provides a lower bound on the variance of an unbiased estimator of a deterministic parameter. To apply the inequality in this context, notice that the function $g$ in the definition of $k$-arbitrage is essentially an estimator of the optimal model $h_\lambda^*(D)$. Hence, for any function $g$ and $\tilde{h} = g(\hat{h}_\lambda^{\delta_1}(D), \hat{h}_\lambda^{\delta_2}(D), \dots, \hat{h}_\lambda^{\delta_k}(D))$, we have:

$$\mathbb{E}\left[\epsilon_s(\tilde{h}, D)\right] \geq \frac{1}{\sum_{j=1}^k \frac{1}{\delta_j}} \tag{6}$$

Suppose that the pricing function $p_{\epsilon_s,\lambda}$ exhibits 1-arbitrage. Then, there must exist parameters $\delta_1, \delta_2$ with $p_{\epsilon_s,\lambda}(\delta_1, D) < p_{\epsilon_s,\lambda}(\delta_2, D)$, and a function $g$ that returns a model $\tilde{h} = g(\hat{h}_\lambda^{\delta_1}(D))$ with

$$\mathbb{E}\left[\epsilon_s(\tilde{h}, D)\right] \leq \mathbb{E}\left[\epsilon_s(\hat{h}_\lambda^{\delta_2}(D), D)\right] = \delta_2.$$

However, Eq. (6) implies that $\mathbb{E}\left[\epsilon_s(\tilde{h}, D)\right] \geq \delta_1$. Thus, we obtain $\delta_1 \leq \delta_2$, which makes condition (2) false.
Next, suppose that the pricing function exhibits $k$-arbitrage for $k \geq 2$. Using the same argument as above, we can show that there exist parameters $\delta_0, \delta_1, \dots, \delta_k$ such that:

1. $\sum_{j=1}^k p_{\epsilon_s,\lambda}(\delta_j, D) < p_{\epsilon_s,\lambda}(\delta_0, D)$; and
2. $1/\delta_0 = \sum_{j=1}^k 1/\delta_j$.

We will show that the above 2 properties imply that condition (1) is false. Indeed, for the sake of contradiction suppose that condition (1) is true, and also that $1/\delta_0 = \sum_{j=1}^k 1/\delta_j$. For $j = 1, \dots, k-1$, let us define $\Delta_j$ such that $1/\Delta_j = \sum_{c=j+1}^k 1/\delta_c$. Observe that $1/\delta_0 = 1/\delta_1 + 1/\Delta_1$, and also $1/\Delta_j = 1/\delta_{j+1} + 1/\Delta_{j+1}$. Then, we can write:

$$\begin{aligned}
p_{\epsilon_s,\lambda}(\delta_0, D) &\leq p_{\epsilon_s,\lambda}(\delta_1, D) + p_{\epsilon_s,\lambda}(\Delta_1, D) \\
&\leq p_{\epsilon_s,\lambda}(\delta_1, D) + p_{\epsilon_s,\lambda}(\delta_2, D) + p_{\epsilon_s,\lambda}(\Delta_2, D) \\
&\leq \cdots \\
&\leq \sum_{j=1}^k p_{\epsilon_s,\lambda}(\delta_j, D)
\end{aligned}$$

This contradicts the first property, and hence condition (1) must indeed be false. □

*of Theorem 7.* We will prove the theorem by showing a reduction from the UNBOUNDED SUBSET-SUM problem. In this problem, we are given as input a set of positive integers $\{w_1, w_2, \dots, w_n\}$, and a positive number $K$. We then want to decide whether there exist non-negative integers $k_i$ such that $\sum_{i=1}^n k_i w_i = K$. In other words, we are asking whether we can achieve sum $K$ using each $w_i$ zero or more times. It is known that UNBOUNDED SUBSET-SUM is $\mathcal{NP}$-hard.

Consider an instance of the UNBOUNDED SUBSET-SUM problem, with positive integers $\{w_1, w_2, \dots, w_n\}$, and a positive number $K$. Without any loss of generality, suppose that $w_1 < w_2 < \cdots < w_n < K$. We now construct an instance for PRICE INTERPOLATION as follows: let $P_j = a_j = w_j$ for $j = 1, \dots, n$, and $a_{n+1} = K$, $P_{n+1} = K + 1/2$. We will prove that there exists a subadditive and monotone function that interpolates the points $(a_j, P_j)$ if and only if there exists no (unbounded) subset sum with value $K$.

$\Rightarrow$ For the first direction, suppose that there exists an unbounded subset sum with value $K$. In other words, there exist positive integers $k_j$ such that $\sum_{j=1}^{n} k_j w_j = K$. For the sake of contradiction, suppose that we can interpolate a subadditive and monotone function $\hat{p}$. Then we have:

$$K + 1/2 = P_{n+1} = \hat{p}(K) = \hat{p}\left(\sum_{j=1}^{n} k_j w_j\right) \leq \sum_{j=1}^{n} k_j \hat{p}(w_j) = \sum_{j=1}^{n} k_j w_j = K$$

which is a contradiction.

$\Rightarrow$ For the reverse direction, suppose that there exists no unbounded subset sum $K$; we will show that we can construct a subadditive and monotone function $f$ that interpolates the $(n+1)$ points. For every $x \geq 0$, define $\mu(x)$ to be the smallest possible unbounded subset sum that is at least $x$. Notice that $\mu(x) \geq x$ for every $x \geq 0$. Then, we define $f(x) = \min\{\mu(x), K + 1/2\}$. It is straightforward to see that $f(x)$ is monotone by construction.

We next show that $f$ interpolates the points. Indeed, for $j = 1, \ldots, n$, we have that $\mu(a_j) = a_j < K + 1/2$ (since $a_j$ by itself gives a sum of $a_j$), and hence $f(a_j) = a_j$. For $j = n+1$, observe that by our starting assumption there is no sum of $K$, and hence $\mu(a_{n+1}) \geq K + 1$, which implies that $f(a_{n+1}) = K + 1/2$.

Finally, we show that $f$ is a subadditive function. Let $x, y \geq 0$. If $\mu(x) \geq K+1$ then, $f(x)+f(y) \geq f(x) = K + 1/2 \geq f(x+y)$. A symmetric argument holds if $\mu(y) \geq K + 1$. Now, suppose that $\mu(x), \mu(y) \leq K$. Then, there exists $k_j, k_j'$ such that $f(x) = \sum_{j=1}^{n} k_j w_j$ and $f(y) = \sum_{j=1}^{n} k_j' w_j$. Now we have:

$$x + y \leq f(x) + f(y) = \sum_{j=1}^{n} k_j w_j + \sum_{j=1}^{n} k_j' w_j = \sum_{j=1}^{n} (k_j + k_j') w_j$$

Hence, if we pick $k_j'' = k_j + k_j'$, we obtain a subset sum that is at least $x + y$. We can then write:

$$f(x+y) \leq \mu(x+y) \leq \sum_{j=1}^{n} (k_j + k_j') w_j = f(x) + f(y).$$

This concludes our proof. □

*of Lemma 8.* Let $\hat{q}$ be a feasible solution to (4). We will show that $\hat{q}$ satisfies the subadditivity condition as well. Let $x, y > 0$. Since $\hat{q}$ satisfies the constraints in (4), we have $\frac{\hat{q}(x)}{x} \geq \frac{\hat{q}(x+y)}{x+y}$ and $\frac{\hat{q}(y)}{y} \geq \frac{\hat{q}(x+y)}{x+y}$. Thus:

$$\hat{q}(x) + \hat{q}(y) \geq \frac{x}{x+y}\hat{q}(x+y) + \frac{y}{x+y}\hat{q}(x+y) = \hat{q}(x+y).$$

This concludes the proof. □

*of Proposition 1.* Consider a feasible solution $\hat{p}$ of (4) of objective value $M$. It is straightforward that $x_i = \hat{p}(a_i)$ is also a feasible solution for (5) with the same objective value.

For the opposite direction, suppose that $\mathbf{x}$ is a feasible solution to problem (5) with value $M$. Without any loss of generality, assume that $a_1 \leq a_2 \leq \cdots \leq a_n$. Let us define $\hat{p}$ to be a piecewise linear function such that:

$$\hat{p}(x) = \begin{cases} \frac{x_{j+1}}{a_{j+1}} x, & x \in [0, a_1] \\ x_j + \frac{x_{j+1} - x_j}{a_{j+1} - a_j}(x - a_j), & x \in [a_j, a_{j+1}] \\ x_n, & x \in [a_n, \infty) \end{cases}$$

It is easy to see that $\hat{p}$ is non-negative, and that for every $i = 1, \ldots, n$ we have $\hat{p}(a_i) = x_i$. Additionally, $\hat{p}$ is monotone, since it is a piecewise linear function where $x_1 \leq x_2 \leq \cdots \leq x_n$. Finally, we show that for any $y \geq x > 0$ we have $\hat{p}(y)/y \leq \hat{p}(x)/x$.
First, assume that $x, y$ are in the same interval $[a, b]$ of the piecewise linear function (which takes values $x_a, x_b$). Since in this interval we have $\hat{p}(x) = x_a + \frac{x_b - x_a}{b-a}(x-a)$, we want to equivalently show that:

$$\frac{x_a}{x} + \frac{x_b - x_a}{b-a}\left(1 - \frac{a}{x}\right) \geq \frac{x_a}{y} + \frac{x_b - x_a}{b-a}\left(1 - \frac{a}{y}\right) \quad \Leftrightarrow \quad \frac{bx_a - ax_b}{x} \geq \frac{bx_a - ax_b}{y} \quad \Leftrightarrow \quad (y-x)(bx_a - ax_b) \geq 0$$

The last inequality holds because $y \geq x$, and also $x_a/a \geq x_b/b$ (which follows from the constraints).
Now, if $x, y$ are not in the same interval, assume that $x$ falls in the $i$-th interval $[a_i, a_{i+1}]$, and $y$ in the $j$-th interval $[a_j, a_{j+1}]$, where $j \geq i$. Then we have:

$$\hat{p}(x)/x \geq \hat{p}(a_{i+1})/a_{i+1} \geq \hat{p}(a_{i+2})/a_{i+2} \geq \cdots \geq \hat{p}(a_j)/a_j \geq \hat{p}(y)/y$$

The first inequality comes from the fact that $x, a_{i+1}$ are in the same interval, the last from the fact that $y, a_j$ are in the same interval, and all the intermediate inequalities from the constraints in (5). $\square$

*of Lemma 9.* Let $\hat{p}$ be a feasible solution of (3). We construct $\hat{q}$ such that for every $x > 0$:

$$\hat{q}(x) = x \cdot \min_{0 < y \leq x} \{\hat{p}(y)/y\}$$

We first show that $\hat{q}$ is a feasible solution of (4). It is easy to see that $\hat{q}$ is always positive. Now consider $0 < x \leq x'$. Then we have:

$$\hat{q}(x)/x = \min_{0<y\leq x}\{\hat{p}(y)/y\} \geq \min_{0<y\leq x'}\{\hat{p}(y)/y\} = \hat{q}(x')/x'$$

To show that $\hat{q}(x) \leq \hat{q}(x')$, define $y_m = \text{argmin}_{0<y\leq x'}\{\hat{p}(y)/y\}$. Now, if $y_m \leq x$, we have $\min_{0<y\leq x}\{\hat{p}(y)/y\} = \min_{0<y\leq x'}\{\hat{p}(y)/y\}$, and the desired result comes from $x \leq x'$. Otherwise, if $y_m > x$, we have:

$$\hat{q}(x) = x \cdot \min_{0<y\leq x}\{\hat{p}(y)/y\} \leq \hat{p}(x) \leq \hat{p}(y_m) = y_m\{\hat{p}(y_m)/y_m\} \leq x' \min_{0<y\leq x'}\{\hat{p}(y)/y\} = \hat{q}(x')$$

Finally, we show that $\hat{p}(x)/2 \leq \hat{q}(x) \leq \hat{p}(x)$ for every $x > 0$. We have already shown that $\hat{q}(x) \leq \hat{p}(x)$. For the first inequality, let as before $y_m = \text{argmin}_{0<y\leq x}\{\hat{p}(y)/y\}$ and define $\Delta = x/y_m \geq 1$. If $\Delta = 1$, then $\hat{q}(x) = \hat{p}(x)$, so the result holds trivially. So, assume that $\Delta > 1$. The key observation is that

$$\hat{p}(x) = \hat{p}(y_m\Delta) \leq \hat{p}(y_m\lceil\Delta\rceil) \leq \lceil\Delta\rceil\hat{p}(y_m)$$

where the second inequality holds from the subadditivity constraint for $\hat{p}$. Thus we have:

$$\hat{q}(x) = x\{\hat{p}(y_m)/y_m\} \geq \frac{\Delta}{\lceil\Delta\rceil}\hat{p}(x) \geq \frac{\Delta}{\Delta+1}\hat{p}(x) > \hat{p}(x)/2$$

where the last inequality follows from the fact that $\Delta > 1$. This concludes the proof. $\square$

*of Proposition 2.* Let $\hat{p}^*$ denote the optimal solution of (3) with optimal value $C_{SA}$, and $\mathbf{x}^*$ the optimal solution of (5) with optimal value $C_{MBP}$. From Proposition 4, there exists a solution $\hat{q}^*$ of (4) that achieves the same value $C_{MBP}$.

From Lemma 8, we obtain that $\hat{q}^*$ is also a solution to (3), and hence it must be that $C_{MBP} \leq C_{SA}$.

Additionally, Lemma 9 tells us that there exists $\tilde{q}$ that is a feasible solution of (4) such that for every $x > 0$, $\hat{p}^*(x)/2 \leq \tilde{q}(x) \leq \hat{p}^*(x)$. If $C'$ is the objective value for $\tilde{q}$, we then have that $C' \leq C_{MBP}$. We next show that $C' \geq C_{SA} + \sum_j T_j(0)/2$.

We first claim that for every $x, i$, we have that $T_i(\tilde{q}(x)) \geq \min\{T_i(\hat{p}^*(x), T_i(\hat{p}^*(x)/2)\}$. Indeed, suppose that this is not true. Then, since $\hat{p}^*(x)/2 \leq \tilde{q}(x) \leq \hat{p}^*(x)$, there exists $\lambda \in [0,1]$ such that $\tilde{q}(x) = \lambda\hat{p}^*(x)/2 + (1-\lambda)\hat{p}^*(x)$. By the concavity of $T_i$, we now have

$$\begin{aligned}T_i(\tilde{q}(x)) &= T_i(\lambda\hat{p}^*(x)/2 + (1-\lambda)\hat{p}^*(x)) \geq \lambda T_i(\hat{p}^*(x)/2) + (1-\lambda)T_i(\hat{p}^*(x)) \\ &> \lambda T_i(\tilde{q}(x)) + (1-\lambda)T_i(\tilde{q}(x)) = T_i(\tilde{q}(x))\end{aligned}$$

which is a contradiction.
Next, we bound $T_i(\hat{p}^*(x)/2)$ as follows using concavity:

$$T_i(\hat{p}^*(x)/2) = T_i\left(\hat{p}^*(x)/2 + 0/2\right) \geq T_i(\hat{p}^*(x))/2 + T_i(0)/2$$

So we can now write:

$$\begin{aligned}T_i(\tilde{q}(x)) &\geq \min\{T_i(\hat{p}^*(x), T_i(\hat{p}^*(x))/2 + T_i(0)/2\} \\ &\geq \min\{T_i(\hat{p}^*(x), T_i(\hat{p}^*(x)) + T_i(0)/2\} \\ &= T_i(\hat{p}^*(x)) + T_i(0)/2\end{aligned}$$

where the last inequality follows from the fact that $T_i$ is non-positive. Finally, we have:

$$C' = \sum_{j=1}^{n} T_j(\tilde{q}(a_j)) \geq \sum_{j=1}^{n} T_j(\hat{p}^*(a_j)) + \sum_{j=1}^{n} T_j(0)/2 = C_{SA} + \sum_{j=1}^{n} T_j(0)/2$$

This concludes the proof. □

*of Proposition 3.* Let $\hat{p}^*$ denote the optimal solution of (3) with optimal value $C_{SA}$, and $\mathbf{x}^*$ the optimal solution of (5) with optimal value $C_{MBP}$. From Proposition 4, there exists a solution $\hat{q}^*$ of (4) that achieves the same value $C_{MBP}$.
From Lemma 8, we obtain that $\hat{q}^*$ is also a solution to (3), and hence it must be that $C_{MBP} \leq C_{SA}$.
Additionally, Lemma 9 tells us that there exists $\tilde{q}$ that is a feasible solution of (4) such that for every $x > 0$, $\hat{p}^*(x)/2 \leq \tilde{q}(x) \leq \hat{p}^*(x)$. If $C'$ is the objective value for $\tilde{q}$, we then have that $C' \leq C_{MBP}$. We next show that $C' \geq C_{SA}/2$. First, notice that $\tilde{q}(x) \leq \hat{p}^*(x)$ implies that for every $j$: $\mathbf{1}_{\tilde{q}(a_j) \leq v_j} \geq \mathbf{1}_{\hat{p}^*(a_j) \leq v_j}$. Now we can write:

$$\begin{aligned}C' &= g_{\text{BV}}(\tilde{q}(a_1), \ldots, \tilde{q}(a_n)) = \sum_{j=1}^{n} b_j \tilde{q}(a_j) \cdot \mathbf{1}_{\tilde{q}(a_j) \leq v_j} \\ &\geq \sum_{j=1}^{n} b_j \tilde{q}(a_j) \cdot \mathbf{1}_{\hat{p}^*(a_j) \leq v_j} \geq \frac{1}{2} \sum_{j=1}^{n} b_j \hat{p}^*(a_j) \cdot \mathbf{1}_{\hat{p}^*(a_j) \leq v_j} = C_{SA}/2\end{aligned}$$

where the last inequality comes from the fact that $\hat{p}^*(x)/2 \leq \tilde{q}(x)$. This concludes the proof. □

# B Dynamic Programming Algorithm for Revenue Optimization

In this section we give the algorithmic details of the dynamic programming for the revenue optimization problem.

# C Brute Force Algorithm for Revenue Maximization with Buyer Valuations

In this section, we provide a brute-force algorithm that solves (3) with objective function $g_{\text{BV}}$.

**Algorithm 1:** Dynamic Programming Algorithm for Revenue Optimization from Buyer Valuation.

**Input** : $v_1 \leq v_2 \leq \cdots \leq v_n, a_1 < a_2 < \cdots < a_n, b_1, b_2, \cdots, b_n \geq 0$
**Output**: Optimal price value $s(1, +\infty)$ and profit $OPT(1, +\infty)$.
% Initialization
$\forall \Delta \in \Delta_{Set} \triangleq \{\frac{v_1}{a_1}, \frac{v_2}{a_2}, \cdots, \frac{v_n}{a_n}, +\infty\}, s_n(n, \Delta) = \min\{v_n, \Delta a_n\}, OPT(n, \Delta) = b_n \cdot s_n(n, \Delta)$
% Main Loop
**for** $k = n - 1$ *to* $1$ **do**
    **for** $\Delta \in \Delta_{Set}$ **do**
        **if** $a_k \Delta \leq v_k$ **then**
            $s_k(k, \Delta) = \Delta a_k,\ s_j(k, \Delta) = s_j(k + 1, \Delta), j > k$
            $OPT(k, \Delta) = b_k \Delta a_k + OPT(k + 1, \Delta)$
        **else**
            $OPT'(k, \Delta) = b_k v_k + OPT(k + 1, v_k/a_k)$
            $OPT''(k, \Delta) = OPT(k + 1, \Delta)$
            **if** $OPT'(k, \Delta) > OPT'(k, \Delta)$ **then**
                $s_k(k, \Delta) = v_k, s_j(k, \Delta) = s_j(k + 1, v_k/a_k), j > k$
                $OPT(k, \Delta) = OPT'(k, \Delta)$
            **else**
                $s_k(k, \Delta) = s_{k+1}(k + 1, \Delta)\frac{a_k}{a_{k+1}}, s_j(k, \Delta) = s_j(k + 1, \Delta), j > k$
                $OPT(k, \Delta) = OPT''(k, \Delta)$
            **end**
        **end**
    **end**
**end**
% NB: In practical implementation, it is not necessary to compute or store $s_j(k, \Delta), j > k$ since it is always equal to $s_j(k + 1, \hat{\Delta})$ for some $\hat{\Delta}$. Thus, we can simply store $\hat{\Delta}$ at $s_{k+1}(k, \Delta)$ instead of copying and storing $s_j(k, \Delta) \forall j > k$. This ensures the algorithm takes only $O(n^2)$ runtime as well as space.

# D Additional Experiments

Now we provide additional experiments on performance of MBP. Figure 11 and Figure 12 present the revenue and affordability gains of MBP compared to other baseline methods. Figure 13 and Figure 14 demonstrate the runtime performance of MBP.

**Algorithm 2:** The Brute Force Algorithm for Revenue Optimization from Buyer Valuation Using Multiple Inter Linear Programming (MILP).

**Input** : $p_1 \le p_2 \le \cdots \le p_K, \delta_1 < \delta_2 < \cdots < \delta_K, f_c^i(\cdot)$
**Output** : Optimal price value $q_1, \cdots, q_K$ and profit *PRO* to the sub-additive constraints.
$OptPrice = \mathbf{0}_{K\times 1}$
$PRO = 0$
$A = \{a : \exists c_k \in N, s.t. \sum_{k=1}^{K} c_k \delta_k = a\}$
Let $a_1 < a_2 < \cdots < a_M$ be all the elements in $A$, and $a_{M+1} = \inf$
**for** $i = 1$ *to* $2^K$ **do**
    $Active = \{k$ : the $k$th bit of $i$ is 1 $\}$
    $Valid = true$
    **for** $j = M$ *to* 1 **do**

$$p^A(a_j) = \min \sum_{w \in Active} PriceUp(w) k_w$$
$$s.t. \sum_{w \in Active} \delta_w k_w \ge a_j$$
$$k_w \in N$$

        **if** $a_j > a_{j+1}$ **then**
            $Valid = false$
            break;
        **end**
    **end**
    **if** *Valid* **then**
        $TempPrice(j) = p^A(\delta_j)$
        $TempPRO = \sum_{k=1}^{K} f_c^k(TempPrice(k))$
        **if** $TempPRO > PRO$ **then**
            $PRO = TempPRO$
            $OptPrice = TempPrice;$
        **end**
    **end**
**end**
$q_j = OptPrice(j)$

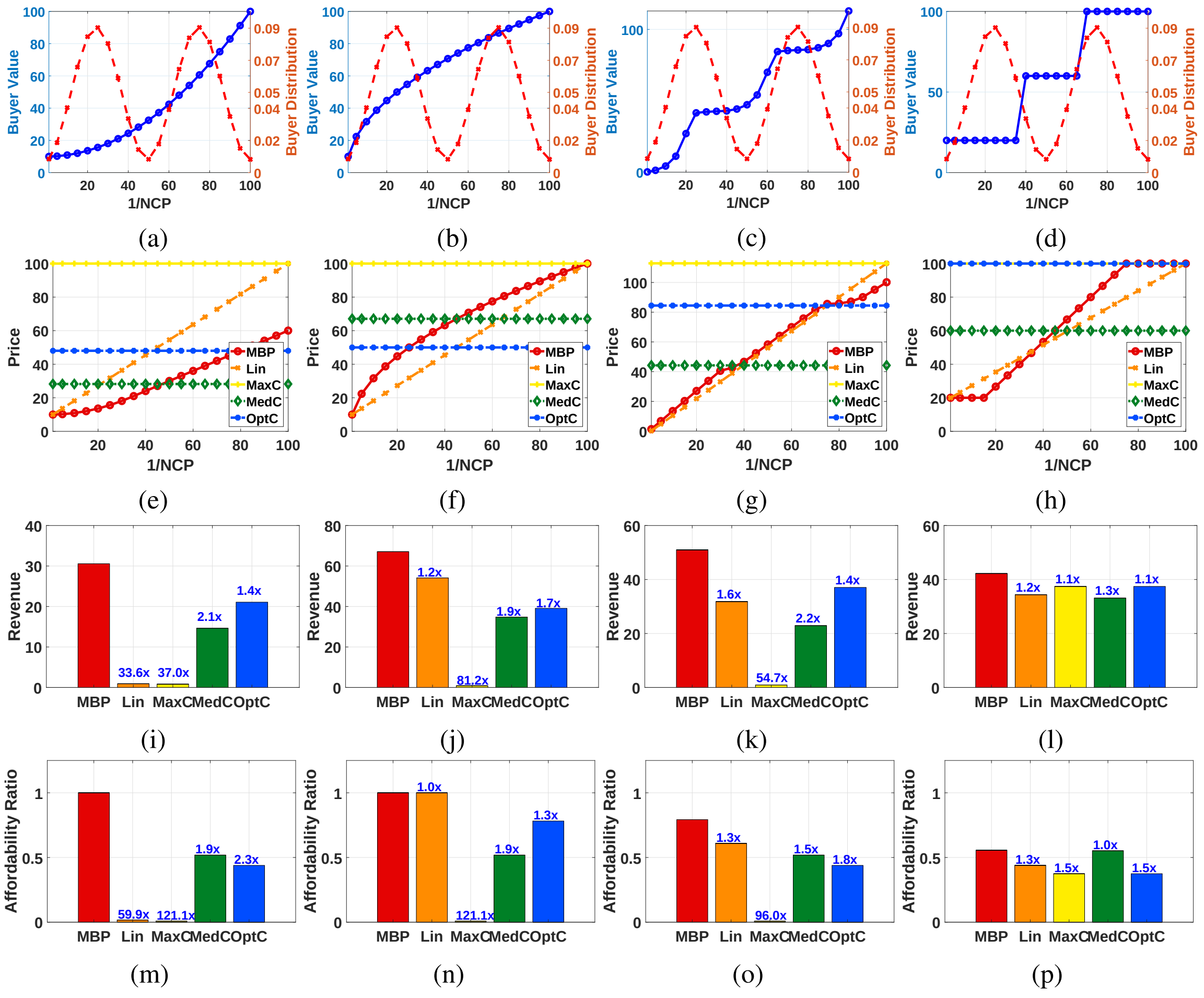

Figure 11: Revenue and Affordability Gain. The buyer distribution is fixed and we vary the buyer value curve.

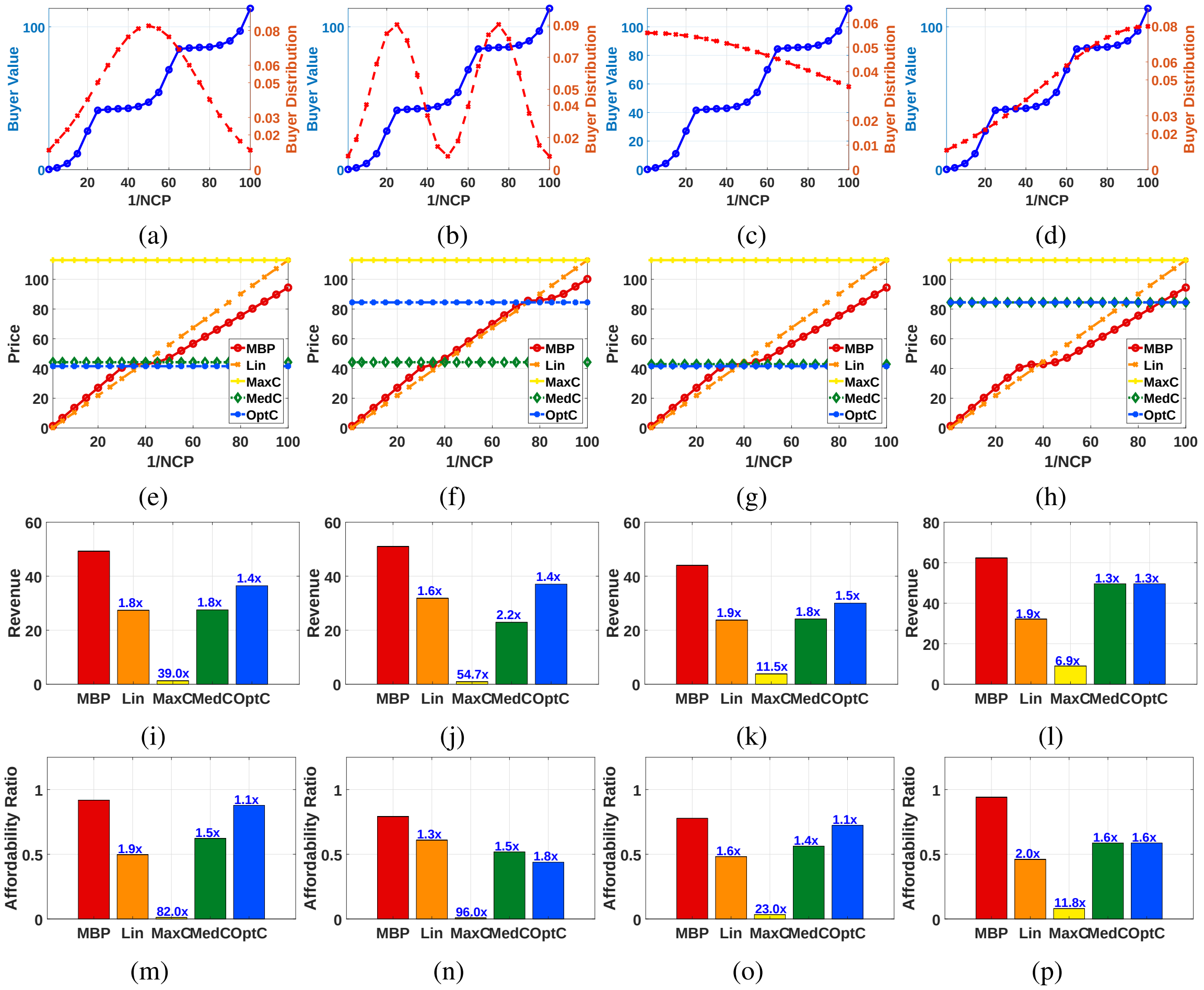

Figure 12: Revenue and Affordability Gain. We fix the buyer valuation and vary the buyer distribution.

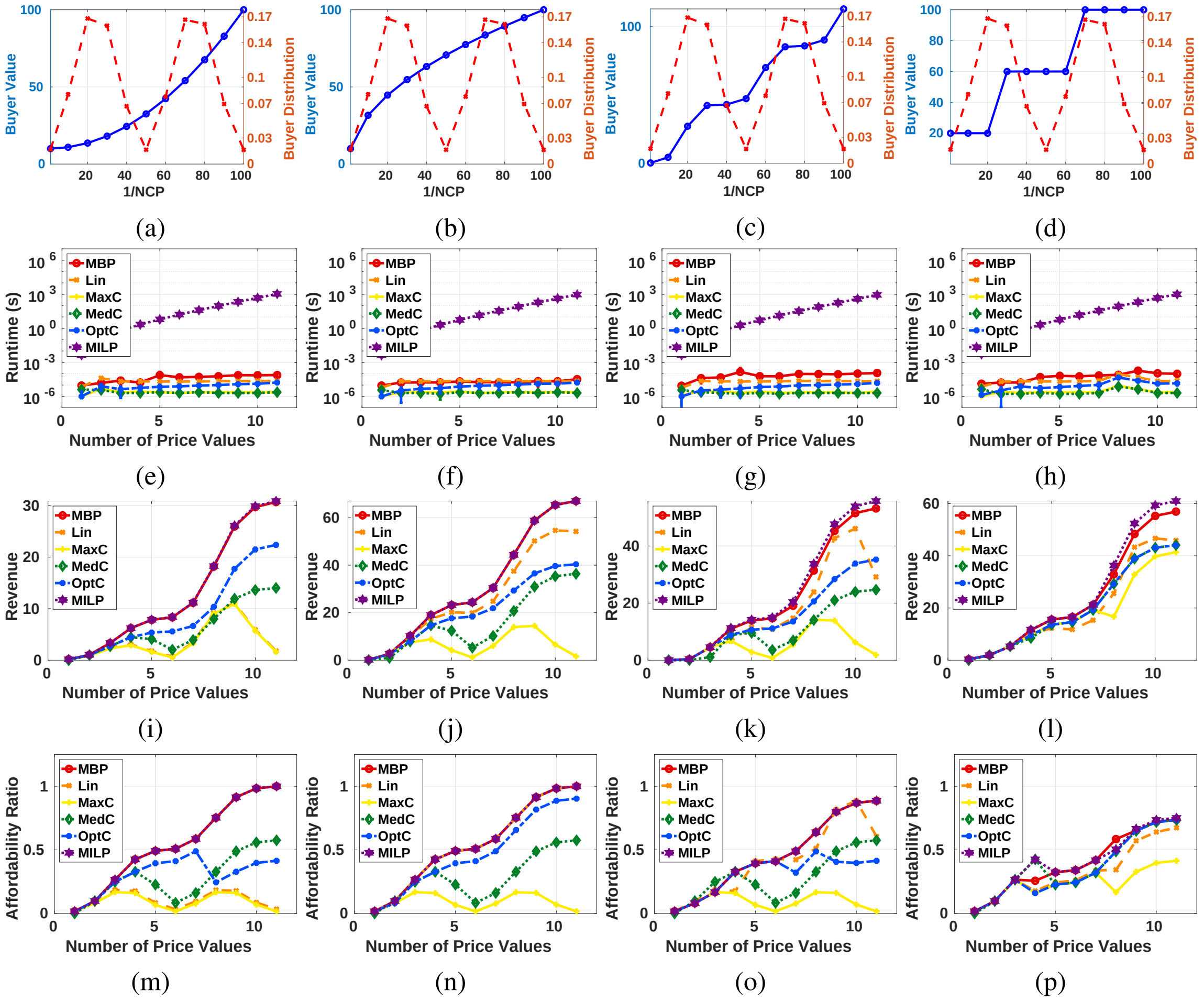

Figure 13: Runtime performance of MBP. We fix the buyer distribution and vary buyer valuation.

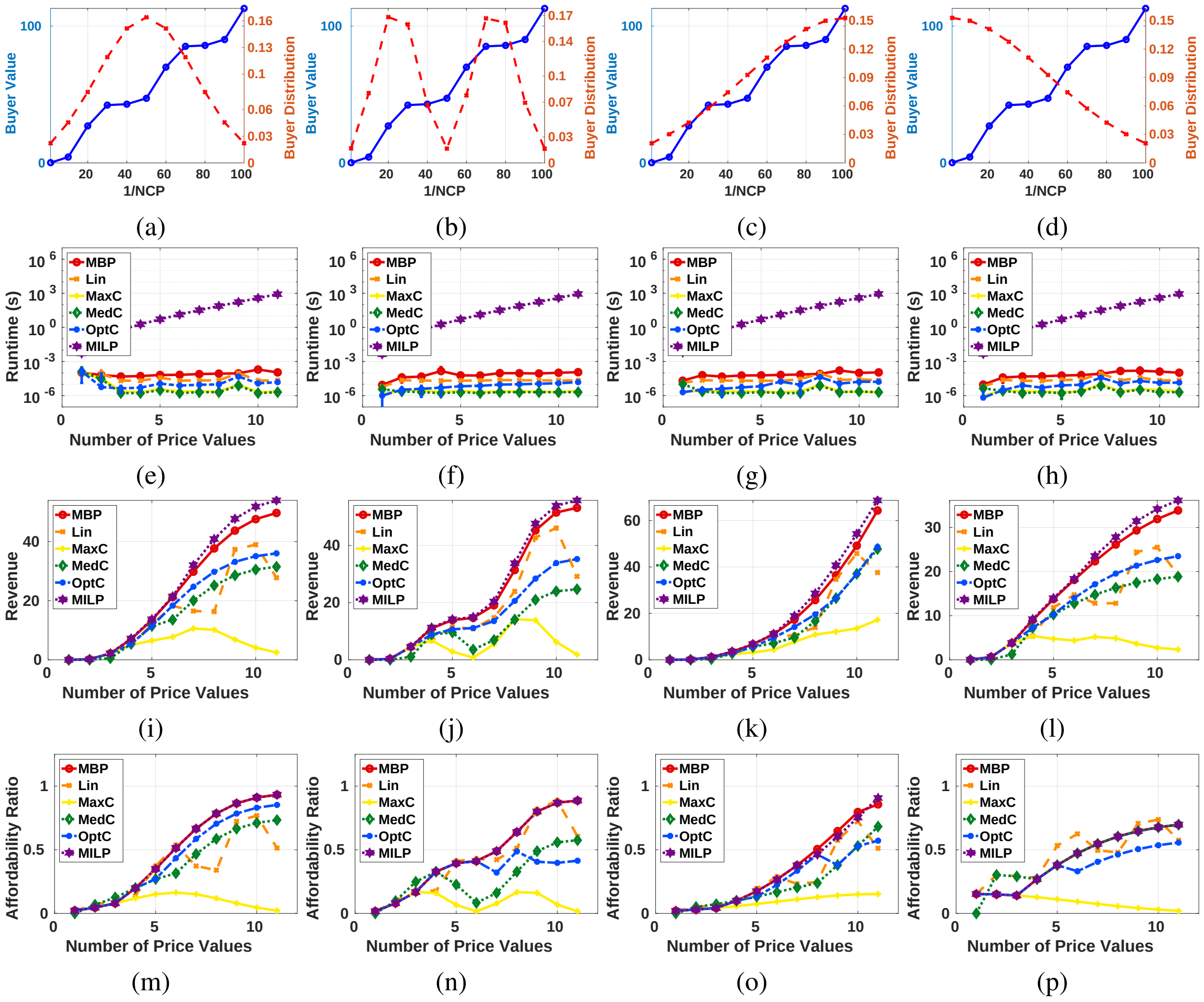

Figure 14: Runtime performance of MBP. We fix buyer value and vary buyer distribution.